\documentclass{article}

\usepackage{arxiv}

\usepackage[utf8]{inputenc} 
\usepackage[T1]{fontenc}    
\usepackage{hyperref}       
\usepackage{url}            
\usepackage{booktabs}       
\usepackage{amsfonts, amsmath}       
\usepackage{nicefrac}       
\usepackage{microtype}      
\usepackage{graphicx}
\usepackage{subfigure}
\usepackage{float}

\usepackage{tikz}
\usetikzlibrary{arrows.meta, positioning, shapes.multipart}

\title{Network Information Enhances Unreliable News Domain Detection}

\author{
    Raphaela Ke{\ss}ler\\
    University of Konstanz, Germany\\
    \And
    Roman David Ventzke\\
    MPI for Dynamics and Self-Organization, Germany\\
    University of Göttingen, Germany\\
    \And
    Viola Priesemann\\
    MPI for Dynamics and Self-Organization, Germany\\
    University of Göttingen, Germany\\
    Complexity Science Hub Vienna, Austria\\
    \And
    Giordano De Marzo \\
    University of Konstanz, Germany\\
    Centro Ricerche Enrico Fermi, Italy\\
    \texttt{giordano.de-marzo@uni-konstanz.de} \\
}

\begin{document}
\maketitle
\begin{abstract}
Content-based detection of unreliable news is increasingly difficult, as low-reliability sources mimic credible journalism and generative AI makes fabricated content harder to flag. We ask whether network structure can improve news reliability classification, taking a domain-level approach that shifts the focus from individual articles to source reliability. From URL-sharing patterns in Telegram chats, we build a statistically validated domain co-sharing network and find assortative mixing by reliability: low-reliability domains group together, as do reliable ones. Exploiting this structure, we compare Graph Neural Networks against network-unaware baselines using both content-aware features (multilingual text embeddings) and content-agnostic features (spreading dynamics). GNNs consistently outperform Multi-Layer Perceptrons on identical features, with GraphSAGE best in both settings (accuracy 0.63 with content, 0.53 without), a 13--14\% relative gain over the network-unaware baseline. Network topology thus systematically improves domain reliability assessment, and remains effective even when content analysis is infeasible.
\end{abstract}

\section*{Introduction}

In the contemporary digital landscape, social networks have fundamentally transformed how information spreads and is consumed. Platforms such as Facebook, YouTube, TikTok and Telegram have become central to the spread of news, especially among younger demographics. The Digital News Report 2024 of Reuters showed that more than a third of the world's population consume news through social media \cite{newman_reuters_2024}, making these platforms primary information sources rather than supplementary channels. Consuming news through online social networks is convenient, cheap and fast but also decentralized and unsupervised \cite{ksiazek_user_2016, khan_fake_2021}. This unprecedented accessibility comes with significant risks, including the formation of echo chambers and filter bubbles, the spread of toxic discourse, and the circulation of low-quality information and fake news. Algorithms designed to maximize engagement often prioritize sensational or emotionally charged content, regardless of its quality, potentially amplifying unreliable information and reinforcing existing biases \cite{wang_viral_2020, cinelli_echo_2021}.

Concerns about information quality have been a prominent public issue since at least the 2016 US elections. In 2017, ``fake news'' was voted word of the year by the Collins Dictionary and ``post-truth'' by the Oxford Dictionary in 2016 \cite{collins-dictionary_collins_2017, oxford-dictionary_word_2016}. The Reuters report revealed that more than half of news consumers are concerned about what is real and what is fake, highlighting the widespread challenge of assessing information quality \cite{newman_reuters_2024}. It is important to stress that information cannot simply be divided into real and fake news. Information quality exists on a spectrum, ranging from high-quality, well-sourced journalism to deliberately fabricated content, with various gradations of reliability, bias, and accuracy in between. Many users have limited skills in evaluating the credibility of online sources, making it difficult to distinguish between high-quality and low-quality information \cite{melchior_systematic_2024}. The COVID-19 pandemic further highlighted these challenges, as social distancing measures, anxiety, and a sense of powerlessness drove people to rely more heavily on digital media for health information, including platforms that facilitate the spread of unreliable information \cite{hornik_association_2021, cinelli_covid-19_2020}.

In the western world, Telegram is known as a prominent platform for misinformation, fake news and low-quality information circulation. In other countries, it is a mainstream or opposition platform~\cite{TeraGram}. In total, with its more than one billion active users, Telegram is one of the largest messaging apps in the world (after WhatsApp, Facebook Messenger, WeChat) \cite{durov_du_2025, ceci_messaging_2025, varela_30_2023}. With its emphasis on privacy and encrypted communication, the platform fosters a sense of security that appeals to users who distrust mainstream authorities. Public chat groups, despite being open to large audiences, maintain a perception of exclusivity, making them ideal spaces for misinformation to spread. Herasimenka et al.\ demonstrated that URLs from lower-quality sources are shared more often on Telegram than content from professional news media, and users who spread lower-quality information were more active content contributors than accounts sharing higher-quality information \cite{herasimenka_misinformation_2023}. After Telegram was blocked in Russia in 2018, founder Pavel Durov called for a ``Digital Resistance,'' emphasizing his commitment to provide a secure and private communication environment and reinforcing the perception of Telegram as a platform for protest activity~\cite{durov_du_2018, wijermars_is_2022}. This further cemented its role as an alternative information ecosystem where non-mainstream or unverified content can circulate anonymously.

Given the vast amount of information shared on social networks and the rise of misinformation, quantifying the reliability of news sources becomes crucial for both researchers and practitioners. However, manual assessment by human experts and fact-checkers, while valuable, faces scalability limitations when confronted with the volume of content circulating on social media platforms. The need for automatic approaches based on machine learning has thus become apparent. Traditional automatic approaches to assessing information reliability have primarily relied on content-based methods, analyzing textual, visual, or audio features of individual pieces of content. However, it is becoming increasingly difficult for content-based models to assess information quality, as low-reliability sources are getting better at mimicking the surface characteristics of high-quality journalism \cite{nguyen_deep_2019}. This problem has been further stressed by the development of powerful Large Language Models (LLMs) that can fabricate realistic, yet fake, news, images or even videos. This challenge has led researchers to explore alternative approaches that examine the structural properties of information networks rather than content features. The fundamental insight underlying this paradigm is that information quality can be inferred from the patterns of how content spreads through social networks, even without analyzing the content itself. Information sources of similar reliability tend to form clusters in sharing networks, as users with particular preferences or biases gravitate toward domains that align with their worldviews \cite{goertzel_belief_1994}. Consequently, Graph Neural Networks (GNNs) have emerged as a promising solution for misinformation detection and news-domain assessment \cite{zhang_fakedetector_2020, carragher_detection_2024}, leveraging their inherent ability to model complex relational structures in social media data. Most research efforts in this area have focused on analyzing the networks formed by the resharing of individual news pieces, particularly examining retweet cascades and propagation trees that emerge when a single news item spreads through social platforms \cite{phan_fake_2023}. Several works, for instance, applied GNNs to information cascades on Twitter, showing that they outperform network-unaware methods \cite{monti_fake_2019, han_graph_2020, gul_advancing_2024}.

Motivated by the developments above, this study investigates whether graph-based learning can improve domain-level reliability classification in Telegram co-sharing networks. We construct an undirected network of news-domain relatedness based on co-sharing patterns observed in Telegram chat groups, leveraging the collective intelligence embedded in user sharing behavior to assess the reliability of entire news domains rather than individual articles. Our analysis reveals that news domains exhibit clear assortative mixing in the co-sharing network, meaning that low-reliability domains tend to be connected to other low-reliability domains, while high-reliability sources cluster together. This structure suggests that users' sharing patterns naturally encode reliability signals that can be exploited for automated assessment. We systematically compare network-aware and network-unaware methods, demonstrating that incorporating network structure consistently improves classification performance. Crucially, this advantage persists even without textual content: using only spreading dynamics, GNNs still outperform their network-unaware counterparts. Our results suggest that the relational patterns in domain co-sharing networks provide valuable complementary signals for reliability assessment, even when rich textual features are available, and become particularly crucial when content-based analysis is limited or unreliable.

\section*{Results}
\begin{figure}[t]
    \centering
    \includegraphics[width=0.95\linewidth]{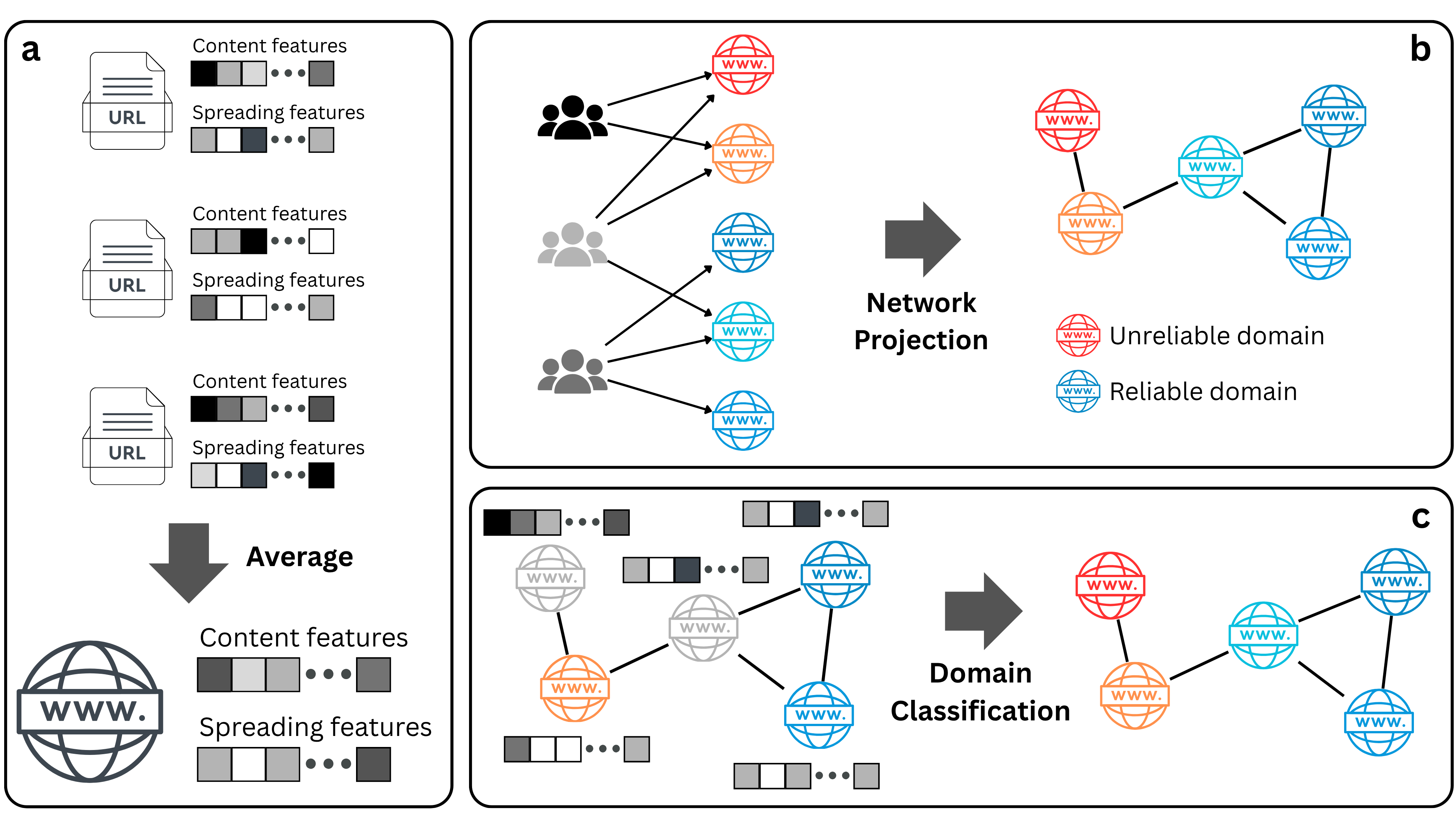}
    \caption{\textbf{Schematic representation of the analysis.} \textbf{a)} Each url is characterized by content features (text embedding) and content agnostic features (spreading features). We then average all urls' features corresponding to the same domain to get content and spreading features at the domain level. \textbf{b)} The sharing of domains on Telegram can be represented by a bipartite network, with one layer containing the chats and the other the domains. Links connect the domains to the chats where they are shared. Moreover each domain is characterized by a reliability score (target variable). The bipartite network is then projected to obtain a similarity network of news domains. \textbf{c)} We finally use the domain-domain network and the content/spreading features to train machine learning models that can predict the quality score of domains.}
    \label{fig:panel}
\end{figure}
\subsection*{News Domains on Telegram and the Reliability Score}
Telegram represents a unique ecosystem for information sharing, structured around public channels and group chats that facilitate large-scale content dissemination. Unlike traditional social media platforms with follower-based networks and algorithmically generated feeds, Telegram operates through broadcast channels where administrators can share content with potentially millions of subscribers, and group chats where members can interact and share external links. This architecture creates natural communities of interest where users aggregate around specific topics, ideologies, or information sources. When users share external news links in these chats, they create implicit signals about source credibility and topic relevance, as the sharing behavior reflects both individual preferences and group dynamics.

Our analysis leverages a comprehensive dataset of Telegram interactions, with about 95\% of the activity concentrated between 2017 and 2023. The data collection employed a snowball sampling methodology, yielding structured information about chat metadata, shared URLs, and temporal sharing patterns. More information on the data collection procedure is reported in the Methods section. For each available URL we construct both content-agnostic and content-aware features (see Fig.~\ref{fig:panel}a). The content-agnostic (or spreading) features, capture the temporal and dissemination dynamics of news sharing without considering textual content. These include, for instance, the magnitude of sharing activity, together with metrics describing the bursty spreading pattern, which is formed by temporally distinct avalanches of coordinated posting activity. Among these is a measure of the burstiness of the temporal structure, which we call ``virality''. Additionally, we include the total number of messages and unique chats that shared the content, serving as proxies for the reach and visibility across the platform. All features are then domain-aggregated and standardized using z-score normalization to ensure comparability across nodes. More details are reported in the Methods. 

For content-aware analysis, we rely on the semantic representations of the actual news content. We systematically scraped textual content from URLs and processed the resulting articles through a multilingual embedding model (paraphrase-multilingual-MiniLM-L12-v2) \cite{hugging-face_sentence-transformersparaphrase-multilingual-minilm-l12-v2_2019}. The embedding of each domain was then defined as the average of these embeddings, i.e. as the average of the embeddings of all the articles from that domain. 

To construct the domain-level network for analysis, we aggregate all URLs with the same domain and represent each domain as a node characterized by content-agnostic or content-aware features. This process is shown in Fig.~\ref{fig:panel}a. We then connected the Telegram chats to news domains, where edges represent the sharing of URLs from specific domains within particular chat groups. As shown in Fig.~\ref{fig:panel}b, this results in a chat-domain bipartite network. It is worth mentioning that in order to maintain focus on news-oriented content, we excluded major social media platforms (Facebook, Instagram, Snapchat, 4Chan) and video platforms (YouTube, Rumble, Bitchute) from our analysis, as their diverse and decentralized content nature would introduce significant classification challenges. 

To establish ground truth labels for the reliability of news reported on a domain, we employ the comprehensive domain rating dataset compiled by Lin et al., which provides credibility assessments for 11,520 news domains \cite{lin_high_2023}. This dataset represents a principled aggregation of six major fact-checking databases through principal component analysis, yielding a unified composite news-reliability score (PC1) that harmonizes quality assessments across different evaluation frameworks and scales. The resulting continuous score reflects the consensus view of multiple fact-checking organizations regarding each domain's credibility and information quality. For our classification task, we discretize this continuous score into three categories: domains with scores below 0.33 are labeled as ``unreliable'' sources, those between 0.33 and 0.66 as ``questionable'' quality, and domains above 0.66 as ``reliable'' sources. The distribution of these quality scores across our dataset reveals a mean of 0.59 and median of 0.63, indicating a slight skew toward higher-quality sources in our Telegram-derived sample, although substantial representation exists across all quality categories. It is worth mentioning, that an average news-reliability score of $0.59$ is substantially lower than other social media platforms like Twitter where the most prevalent domains are concentrated between $0.6$ and $1.0$ \cite{TeraGram}, underlining the high prevalence of low-reliability information that can be found on Telegram.

\begin{figure}[t]
    \centering
    \includegraphics[width=0.5\linewidth]{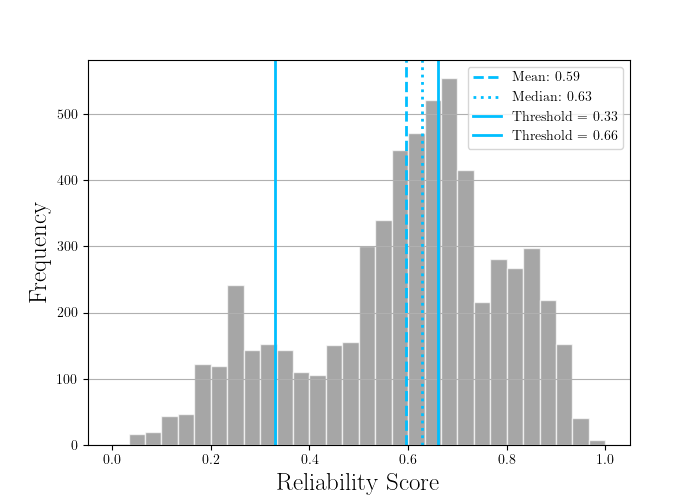}
    \caption{\textbf{Distribution of news-domain reliability on Telegram.} Histogram of
the composite reliability score (PC1) of the news links shared across Telegram chats,
obtained by aggregating six major fact-checking databases through principal component
analysis; higher values denote more reliable sources. The dashed and dotted vertical
lines mark the distribution mean ($0.59$) and median ($0.63$), while the two solid
lines at $0.33$ and $0.66$ indicate the thresholds used to discretize domains into the
\emph{unreliable} ($<0.33$), \emph{questionable} ($0.33$--$0.66$), and \emph{reliable}
($>0.66$) classes.}
    \label{fig:misinformation_hist}
\end{figure}

\subsection*{The Network Structure}
Our starting bipartite network comprises 24,658 Telegram chats and 6,106 domains (4,884 training, 1,222 test), connected by 1,519,640 chat-domain sharing relationships in the original bipartite structure. Rather than directly using this bipartite network, we focus on the relationships between news domains based on their co-sharing patterns in Telegram chats. The idea is that domains that are frequently shared together within the same chats likely serve similar audiences, cover related topics, and, most important, are characterized by comparable credibility characteristics. As schematized in Fig.~\ref{fig:panel}b, we thus construct a domain-domain network through projection of the original bipartite chat-domain structure. This projection allows us to capture implicit similarities between domains. The resulting monopartite network enables direct analysis of domain relationships and provides the foundation for graph-based quality assessment approaches.

To construct a meaningful domain co-occurrence network, we employ the Bipartite Configuration Model (BiCM) to validate the statistical significance of domain relationships \cite{Saracco_2015, vallarano_fast_2021, Saracco_2017, BICM_Github}. The BiCM generates randomized versions of the bipartite network while preserving (on average) the degree sequences of both chat and domain nodes, enabling identification of domain co-occurrences that exceed what would be expected by chance given the popularity distributions of both chats and domains \cite{Saracco_2015, vallarano_fast_2021}. This validation process filters out spurious connections that arise merely from the high activity of popular domains or chats, yielding a statistically validated backbone that captures genuine structural relationships in the data \cite{Saracco_2017}.

\begin{figure}[!ht]
  \centering
  \subfigure[]{%
    \includegraphics[width=0.45\linewidth]{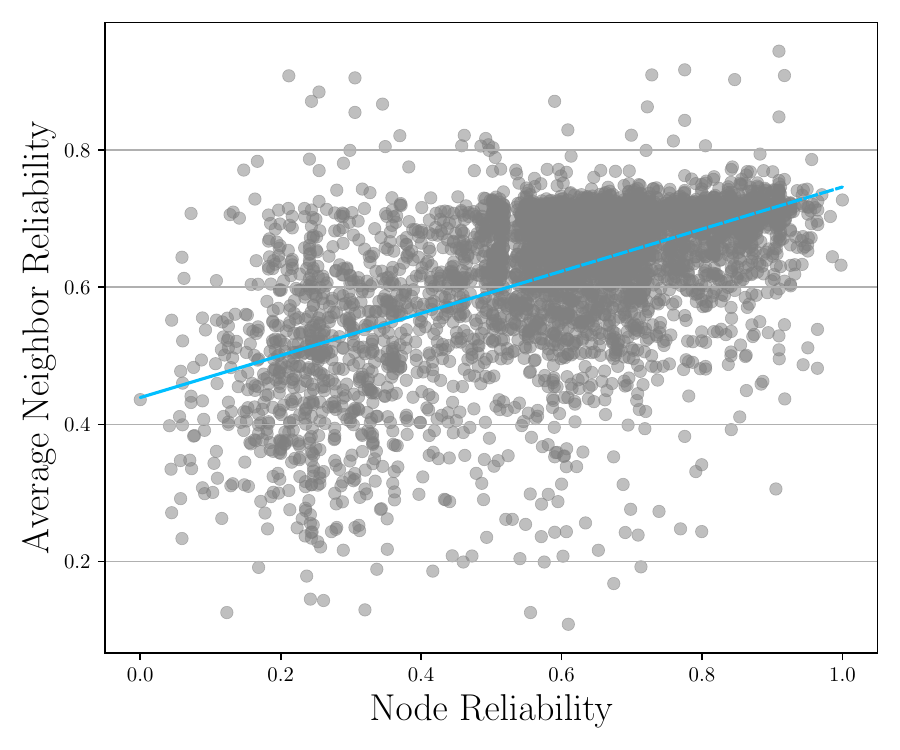}%
    \label{fig:neighbor_assortativity}}
  \hfill
  \subfigure[]{%
    \includegraphics[width=0.5\linewidth]{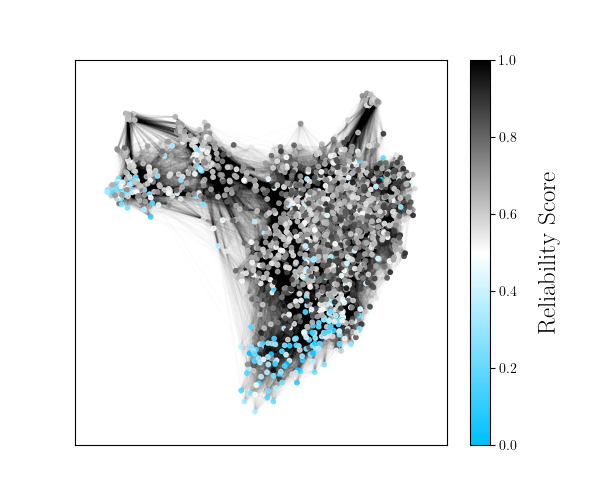}%
    \label{fig:val_network}}

  \caption{\textbf{Assortative mixing of reliability in the domain co-sharing network.}
\textbf{a)} For each domain, the average reliability score of its network neighbors is
plotted against the domain's own reliability score. The positive trend (blue regression
line) reflects assortative mixing: domains tend to be co-shared with others of comparable
reliability, quantified by a reliability assortativity coefficient of $0.22$.
\textbf{b)} Visualization of the statistically validated monopartite domain--domain
network, restricted to the top $50\%$ of nodes by degree for readability. Node color
encodes the reliability score, from low-reliability (light blue) to high-reliability
(dark) sources. Low- and high-reliability domains segregate into distinct clusters,
confirming that the network topology carries substantial information about domain
reliability, independently of node content features.}
  \label{fig:combined}
\end{figure}

The resulting validated domain network is reported in table \ref{tab:network_statistics}. Of the 4,884 domains in the training set, 93.7\% survive the validation and constitute the nodes of the network, while the remaining 310 domains share no co-occurrences beyond chance expectation and are pruned. The network contains 4,574 statistically validated nodes with an average degree of 153.43, indicating that domains are, on average, significantly co-shared with approximately 153 other domains. Despite this substantial connectivity, the overall network density remains relatively sparse at 0.03. Most importantly for our reliability assessment task, the network demonstrates assortative mixing with respect to the reliability score, as measured by a reliability assortativity coefficient of 0.22. This indicates that domains of similar reliability scores tend to be connected to each other at rates significantly higher than would be expected by chance. Statistical significance of the reliability assortativity was assessed with a permutation test (5,000 permutations) in which node reliability scores were shuffled while the network structure was held fixed. The observed coefficient (r = 0.22) exceeded all permuted values (p < 0.001, z = 149.4), confirming that domains of similar quality co-occur significantly more often than expected under the null model. Fig.~\ref{fig:neighbor_assortativity} illustrates this pattern clearly, plotting the relationship between individual domain quality scores and the average quality scores of their network neighbors. Fig.~\ref{fig:val_network} shows the top 50\% of nodes by degree in the validated network; node color represents the reliability score of each domain. Unreliable sources tend to form clusters that are well separated from the reliable sources. This cluster structure and the presence of assortativity, suggest that users' sharing behaviors naturally encode reliability signals, as domains that serve similar information ecosystems or audience segments tend to be co-shared and thus connected in our network. As a consequence we expect the network structure to encode useful information that can complement the node features in detecting the credibility of domains. The final stage of the pipeline then involves utilizing the constructed network, together with the extracted domain-level features, as input to machine learning models designed to predict the reliability category of news domains. (Fig.~\ref{fig:panel}c).

\subsection*{Models}
We apply several GNN architectures, including the Graph Convolutional Network (GCN) \cite{kipf_semi-supervised_2017}, Graph Attention Network (GAT) \cite{velickovic_graph_2018} and GraphSAGE \cite{hamilton_inductive_2017} to our network of domains. The input consists of a graph $G$ along with node features $X$, which are split into content and context components. The output is instead the predicted node labels $\hat{y}$, with three different classes (Unreliable=0-0.33, Questionable=0.33-0.66, Reliable=0.66-1).

The GNN architectures were implemented using PyTorch Geometric \cite{fey2019fast}, an extension library built on PyTorch \cite{paszke2019pytorch} that supports scalable message-passing paradigms for graph-based learning tasks. The GNN architectures are summarized in Table \ref{tab:gnn-architectures} and differ in depth, dropout regularization, and message-passing mechanisms. All models are trained using the Adam optimizer with a negative log-likelihood loss, and hyperparameters are tuned per model to optimize classification accuracy. The selected hyperparameters and their respective candidate values are shown in Table \ref{tab:param_grid}. As baseline model we implement a Multi-Layer Perceptron (MLP) as described in the Methods section.

\begin{table}[h!]
\centering
\caption{Summary of Graph Neural Network Architectures}
\label{tab:gnn-architectures}
\begin{tabular}{lcllc}
\toprule
\textbf{Model} & \textbf{Layer Count} & \textbf{Layer Type} & \textbf{Activation} & \textbf{Dropout} \\
\midrule
GCN & 3 & GCNConv & ReLU & Yes \\
GAT & 3 & GATConv& ELU & Yes \\
GraphSAGE & 3 & SAGEConv & ReLU & Yes \\
MLP & 3 & Linear & ReLU & Yes \\
\bottomrule
\end{tabular}
\end{table}

\subsection*{Evaluation}
Table \ref{tab:combined_results} summarizes the results for the three GNN architectures alongside a MLP baseline. All confusion matrices and learning curves for the associated models can be viewed in the supplementary material. The GraphSAGE model achieved the best performance with an accuracy and F1-score of 0.63, outperforming all other GNNs as well as the MLP baseline. This result is interesting given that GraphSAGE was designed for inductive learning, allowing the model to generalize better across unseen nodes \cite{hamilton_inductive_2017}. The GraphSAGE model can successfully classify 101 out of the 171 news domains as unreliable. The GAT model, with its neighbourhood aggregation by attention mechanisms, performed similarly to GCN (accuracy=0.57). With an F1 score and accuracy of 0.55, the MLP baseline performed about 8 percentage points worse than the GraphSAGE model. In fake news detection, it is crucial to minimize false negative cases, as incorrectly labeling fake news as true can facilitate the spread of misinformation and reduce the system’s credibility. The simple Multi-Layer Perceptron (MLP) model generates more false negatives than the GraphSAGE model, as it is less effective at capturing the relational dependencies between news items and their associated sources in the Telegram chat group.

These results support the hypothesis that GNNs, particularly those designed for inductive data, can outperform simpler models like Multi-Layer-Perceptron in misinformation detection tasks.\\

\begin{table}[ht]
\caption{Performance comparison of GNN models with and without content features}
\label{tab:combined_results}
\centering
\begin{tabular}{lcccc}
\toprule
\textbf{Model} & \textbf{Accuracy} & \textbf{Precision} & \textbf{Recall} & \textbf{F1-Score} \\
\midrule
\multicolumn{5}{c}{\textbf{With Content Features}} \\
\midrule
GCN & 0.57 & 0.57 & 0.57 & 0.57 \\
GAT & 0.57 & 0.58 & 0.57 & 0.56 \\
GraphSAGE & \textbf{0.63} & \textbf{0.63} & \textbf{0.63} & \textbf{0.63} \\
MLP Baseline & 0.55 & 0.56 & 0.55 & 0.55 \\
\midrule
\multicolumn{5}{c}{\textbf{Without Content Features}} \\
\midrule
GCN & 0.50 & 0.53 & 0.50 & 0.47 \\
GAT & 0.50 & 0.52 & 0.50 & 0.48 \\
GraphSAGE & \textbf{0.53} & \textbf{0.54} & \textbf{0.53} & \textbf{0.52} \\
MLP Baseline & 0.47 & 0.43 & 0.47 & 0.43 \\
\midrule
\midrule
Random Baseline & 0.36 & 0.36 & 0.36 & 0.36 \\
Simple Network Classifier & 0.38 & 0.39 & 0.38 & 0.38 \\
\bottomrule
\end{tabular}
\end{table}

To better understand the importance of textual features, the same set of GNN models was tested without semantic content embeddings, relying only on metadata such as virality, avalanche frequency, and chat interactions. Results are presented in Table \ref{tab:combined_results}. As in the content-based setting, GraphSAGE was the top performer, achieving an accuracy of 0.53 and an F1-score of 0.52. After 100 epochs, the model can successfully classify 42 domains of the test set as unreliable. The GraphSAGE models show a difference of about 10 percentage points in accuracy between the content and content-agnostic approaches. Interestingly, the performance drop between content-based and content-agnostic variants is relatively small, suggesting that topological and behavioral patterns in URL sharing encode substantial signals for classification. This supports the growing view in misinformation research that content-agnostic features, such as propagation patterns, can be as informative as linguistic cues \cite{monti_fake_2019, zhou2019network}. The MLP baseline, which lacks any graph awareness, scored an accuracy of 0.47 and an F1-score of 0.43, reinforcing the added value of GNN architectures even without content features. 

In sum, while content-based models provide the best performance, the content-agnostic GNN achieved almost competitive results, offering a language-independent, resource efficient and interpretable solution for misinformation detection on platforms like Telegram.

\section*{Discussion}
Content-based detection of misinformation is increasingly strained, as low-reliability sources mimic the stylistic and structural features of credible journalism and generative AI makes fabricated text, images, and video harder to flag. 
On the other hand, network-based approaches offer a complementary route, exploiting the structural and relational properties of information sharing rather than content alone. The underlying idea is that how information propagates through social networks often reveals more about a source's reliability than its content: users tend to consume and share from sources aligned with their perspectives, producing clustering patterns that reflect the reliability of the underlying domains. These structures emerge organically from collective user behavior and encode reliability signals that can be exploited even when content analysis is unreliable or resource-intensive. Motivated by these ideas, in this work, we used Graph Neural Networks (GNNs) to investigate whether network structure improves the automated assessment of news-domain reliability, and found that it does so consistently.

Concretely, we consider news domains shared on Telegram, a prominent platform for the circulation of unreliable news. Rather than analyzing the propagation cascades of individual news items, we constructed a domain co-sharing network from URL-sharing patterns across Telegram chat groups, with about 95\% of the activity concentrated between 2017 and 2023. We represented this ecosystem as a bipartite network connecting chat groups to news domains, which we then projected into a domain similarity network where edges reflect the statistically validated frequency of co-occurrence in the same chats. Each domain node was characterized by both content-aware features, derived from multilingual embeddings of scraped article text, and content-agnostic features capturing spreading dynamics such as virality, avalanche patterns, and reach metrics. To measure domain reliability, we exploited a domain reliability dataset that aggregates assessments from six major fact-checking organizations into a unified reliability score, which we discretized into three categories: unreliable, questionable, and reliable. We then trained and evaluated multiple Graph Neural Network architectures (Graph Convolutional Networks, Graph Attention Networks and GraphSAGE) against a network-unaware baseline consisting of Multi-Layer Perceptrons operating on the same feature sets.

As a first result, the projected domain co-sharing network exhibits clear assortative mixing with respect to reliability: domains tend to be more likely connected to other domains with similar reliability score. This assortativity manifests as a clustering pattern in the network, with low-reliability sources forming distinct groups. It demonstrates that the topology of the co-sharing network itself carries substantial information about domain reliability, independent of content features. Exploiting this structure, Graph Neural Networks consistently outperform network-unaware models for domain reliability classification, whether or not content features are available. In the content-rich setting, GraphSAGE achieved 63\% accuracy and a 63\% F1-score, compared to 55\% accuracy and 55\% F1-score for the MLP baseline, an improvement of roughly 8 percentage points, or about 14\% in relative terms. The gap persisted in the content-agnostic setting, where only spreading dynamics were available: GraphSAGE reached 53\% accuracy versus 47\% for the MLP, again about 6 percentage points (13\% relative). This advantage over the network-unaware baseline holds across all tested architectures and is largest for GraphSAGE, indicating that the benefit derives from incorporating network structure rather than from specific architectural choices. Even when rich textual embeddings are available, GNNs leverage the structural context of neighboring domains to refine their predictions. This has a practical implication for unreliable-news detection systems: effective reliability assessment can be achieved through metadata and network analysis even when content scraping is technically infeasible, legally restricted, or computationally prohibitive.

While these findings demonstrate the promise of network-based reliability assessment, our work has limitations worth noting. First, the analysis is restricted to Telegram and operates at the domain level rather than the granularity of individual URLs. Although Telegram is a highly relevant platform given its prominence in the circulation of low-reliability information, this focus means the structural properties and sharing dynamics we observe may not directly generalize to other ecosystems with different architectures, user bases, and moderation policies. Moreover, domain-level aggregation, while practical and interpretable, obscures within-domain heterogeneity: not all content from a domain maintains uniform reliability, and individual articles may deviate substantially from the domain's average credibility. Additionally, a recent review of best practices for domain reliability scores such as NewsGuard recommends continuous rather than discrete reliable/unreliable ratings, for finer resolution and greater temporal stability~\cite{Luhring2025}. Future work could extend the approach to other platforms and to finer granularities that preserve more information about content variation.

Second, we emphasize that the primary objective of this work was not to achieve state-of-the-art performance, but to establish the principle that incorporating network structure systematically improves reliability assessment relative to network-unaware approaches. Our results clearly support this: across all experimental settings, GNN models consistently outperform baselines operating on identical features but ignoring network topology. The modest absolute performance levels nonetheless leave substantial room for improvement. We used a single, relatively lightweight embedding model (paraphrase-multilingual-MiniLM-L12-v2) for content representation, and tested standard GNN architectures without extensive hyperparameter optimization or architectural innovation. Stronger performance could likely be obtained through several enhancements: more powerful large-scale multilingual language models for content embedding; more sophisticated architectures, such as heterogeneous graph networks that explicitly model the multi-relational chat--domain ecosystem; larger and better curated news datasets; and temporal graph networks that capture how sharing patterns and domain reliability evolve over time. Developing such models is a promising direction for future work, building on the foundational insight established here that network structure provides valuable and complementary information for unreliable-news detection.

\section*{Methods}
\subsection*{Data Set}
We used a large Telegram data-set which has been obtained by crawling public groups chats and channels (we refer to both as 'chats') on Telegram through a snowball-sampling inspired algorithm described in \cite{mohr2023inference,TeraGram}. Starting with popular seed channels that were downloaded using the Telegram API,  we iteratively followed forwarded messages - which act as links in the Telegram network - to find new chats. These new found chats were then added into a priority driven download queue. The priority of a chat in the queue was determined by its relatedness to the target topic - in this case the COVID-19 pandemic, represented by a multi-lingual list of topical key words. Since the content of a new chat is unknown pre-download, its relatedness to the target topic was estimated based on available information (title and description of the chat) and the content of downloaded neighboring chats (chats that forwarded messages from the new chat). The result is a guided sampling procedure that explores the public chat network while preferably downloading chats that have the target key-words. We are confident that despite that priority sampling, the data set is very general, because prioritization flattened after about downloading 10 \% of the total data set, leading to the inclusion of a majority of chats that are not related to COVID-19. A detailed description of the algorithm and data-set can be found in \cite{mohr2023inference}. A second data set \cite{TeraGram} was obtained without the COVID-19 topic guiding mechanism, and we used it to check whether the COVID-topic sampling did impact the misinformation rate.  We found that the overall prevalence of misinformation between the guided sample used here and the unguided sample in the newer data-set does not change significantly when measured according to the domain-reliability score (compare Fig. \ref{fig:misinformation_hist} and Fig. 6 in \cite{TeraGram}). This occurs despite the high prevalence of health-related misinformation on Telegram \cite{Rieskamp2024, Wehrli2025}, which is in-line with the observation that the chat priority is important early during a sampling run, but decays quickly once larger fractions of the public chat network are explored \cite{mohr2023inference}. 
The data can be made available upon reasonable request. 

\subsection*{Feature Extraction}
For the downloaded chat sample ($128$k chats), the most prevalent language of each chat was classified using text samples and the langdetect library \cite{langdetect,langdetectport}. We restrict our analysis to English speaking chats (24,658 chats with $\approx 400$M messages), since the employed domain reliability measure is more complete and consistent for English speaking news domains \cite{Luhring2025}. We identified web URL occurrences within the texts of messages (total of $322$M across $172$k domains) and built a bi-partite network of URLs and chats with weighted connections according to the sharing frequency of a URL in each chat. For each URL, content-agnostic features of the spreading dynamics in the chat network are then computed. To describe the spreading dynamics, we decomposed the time-series of posts that contain the target URL into distinct, so called, avalanches (bursts of activity) to capture the bursty nature of the spread of online content \cite{Notarmuzi2022}. This was achieved by clustering all posts that occur without a break of more than $\Delta t = 1\;\mathrm{h}$ into one avalanche. Features describing the spreading dynamics of a URL are:\begin{itemize}
    \item the total number of chats that shared the URL and the total number of posts containing the URL
    \item the total number of avalanches $n$
    \item the average size (number of sharing events) of avalanches $\frac{1}{n}\sum_{i=1}^n s_i$
    \item the maximal size of an avalanche $\mathrm{max}(s_1,..,s_n)$
    \item the virality of the avalanche sequence $\frac{\mathrm{max}(s_1,..,s_n)}{\sqrt{\sum_{i=1}^{n}s_i^2}}$: a measure for the relative size of the largest avalanche in the temporal sequence of avalanches $s_1,..,s_n$ that quantifies how "bursty" the temporal spreading pattern is.
    \item the peak sharing activity: maximal number of posts that share the URL within time $T$ for a range of time-spans $T \in \{1\,\mathrm{h},6\;\mathrm{h},1 \; \mathrm{d}, 3 \; \mathrm{d}, 14 \; \mathrm{d}\}$
\end{itemize}
Further, content-based features were computed from the text extracted from each URL. For each domain, we retained the 1000 most frequently forwarded URLs and scraped the first 1000 characters of each article using Selenium. We embed the text using the multilingual SBERT model paraphrase-multilingual-MiniLM-L12-v2 and treat the resulting vector as a content feature \cite{hugging-face_sentence-transformersparaphrase-multilingual-minilm-l12-v2_2019} \footnote{The model supports over 50 languages and maps text into a 384-dimensional embedding space. It is trained on large-scale parallel sentence pairs from OPUS resources (e.g., OPUS-100, Europarl, and OpenSubtitles) \cite{muennighoff_mteb_2023}}. The data was then coarse-grained to the level of domains, by assigning averaged features of all respective URLs to each of the domains. Here we restrict ourselves to the 6106 domains that have a reliability score, featured in the Lin et al. \cite{lin_high_2023} data-set, and additionally exclude social-media sites and video platforms, since content on these platforms can have a very high variability which does not suit our domain-level model \footnote{We excluded the platforms Facebook, Instagram, Snapchat, 4Chan, YouTube, Rumble and Bitchute}. The coarse-grained bi-partite sharing network of domains and chats was then constructed by aggregating over all URLs of the respective domains. Projection of this bi-partite network onto the domains yields the domain co-occurrence network. Finally, to remove insignificant connections in the graph, we employed the Bipartite Configuration Model \cite{BICM_Github,Saracco_2015,Saracco_2017}, and pruned connections from the graph that are likely to be generated through random chance.

\subsection*{Data Preprocessing}

The dataset was partitioned into three reliability-based classes derived from the ground truth reliability score (PC1) as described in the Results section and summarized in Table \ref{tab:label_distribution} \cite{lin_high_2023}. Class 0 corresponds to unreliable (PC1 range: 0–0.33), Class 1 to questionable information (PC1 range: 0.33–0.66), and Class 2 to reliable information (PC1 range: 0.66–1). The resulting label distribution is relatively balanced, with a slight dominance of the questionable and reliable categories. Data were split into training and test subsets using a fixed 80/20 ratio, maintaining approximately equal class proportions across both sets.

\begin{table}[h!]
\centering
\caption{Label distribution (PC1 bins) across the train/test split. The Validated column reports the training nodes surviving BiCM validation, described in the Feature Extraction section.}
\label{tab:label_distribution}
\begin{tabular}{ccl ccc}
\toprule
 & & & \multicolumn{2}{c}{\textbf{Train}} & \textbf{Test} \\
\cmidrule(lr){4-5} \cmidrule(lr){6-6}
\textbf{Class} & \textbf{PC1 Range} & \textbf{Description} & \textbf{Raw} & \textbf{Validated} & \textbf{Raw} \\
\midrule
0 & $0 - 0.33$ & Unreliable Information & 720 (14.7\%) & 643 (14.1\%) & 171 (14.0\%) \\
1 & $0.33 - 0.66$ & Questionable Information & 2125 (43.5\%) & 1971 (43.1\%) & 549 (44.9\%) \\
2 & $0.66 - 1$ & Reliable Information & 2039 (41.8\%) & 1960 (42.9\%) & 502 (41.1\%) \\
\midrule
\multicolumn{3}{l}{\textbf{Total}} & 4884 & 4574 & 1222 \\
\bottomrule
\end{tabular}
\end{table}

\subsection*{Network Properties}

The summary table \ref{tab:network_statistics} gives an overview of the network measures of the validated monopartite projection of the chat-domain network. The average degree of 153.43 indicates that, on average, each domain is linked to about 153 of others, suggesting frequent sharing patterns and extensive information overlap among chats. The degree range, spanning from 1 to 935, highlights a highly heterogeneous structure with a few hubs playing a central role in content dissemination \cite{furht_social_2010}. The network density describes the ratio of actual connections to all possible connections in a network and thus indicates how close the network is to be fully connected. Despite this variability, the network density of 0.03 remains relatively low, consistent with typical real-world social networks that are sparse at scale. The degree assortativity coefficient of 0.39 suggests a moderate tendency for nodes to connect with others of similar degree, implying that well-connected domains preferentially link with other highly connected domains.

\begin{table}[h!]
\centering
\caption{Network Summary Statistics.}
\label{tab:network_statistics}
\begin{tabular}{lc}
\toprule
Measurement & Value \\
\midrule
Nodes & 4574\\
Edges & 350883 \\
Average Degree & 153.43 \\
Degree Range & 1 -- 935 \\
Density & 0.03 \\
Degree Assortativity & 0.39 \\
Reliability Assortativity & 0.22 \\
\bottomrule
\end{tabular}
\end{table}

\subsection*{Random and Baseline Model}

This reliability assortativity enables a straightforward density-based approach for domain reliability prediction, where the predicted reliability of any domain can be estimated as the average reliability score of its network neighbors. This neighborhood averaging method serves as one of our three baseline models, representing the intuitive hypothesis that a domain's reliability can be inferred from the company it keeps in the sharing network. By comparing GNN performance against this density-based method, we can assess whether explicit modeling of network structure through neural architectures provides meaningful improvements over simple neighborhood aggregation for domain reliability assessment.
Overall we compare GNN models against three baselines: (i) a random classifier as a chance-level reference, (ii) the neighbor-averaging network classifier as a non-parametric structural baseline, and (iii) an MLP baseline that uses the same input features but does not explicitly model graph structure. We used a MLP with three fully connected layers, the first layer maps the input features to a hidden representation, followed by a second and third hidden layer of equal dimensionality. All hidden layers use the ReLU activation function to introduce non-linearity and enable the model to capture complex feature interactions. This architecture is intentionally simple and does not incorporate any graph information, making it well-suited to serve as a benchmark for evaluating the added value of GNN-based models.

\subsection*{Model Architecture}

The models implemented in this study utilize three types of graph convolution layers: GCNConv, GATConv, and SAGEConv, all from the PyTorch Geometric library \cite{pyg-team_torch_geometricnnconvgcnconv_2025}. The GCNConv layer applies a first-order approximation of spectral graph convolutions, using the normalized graph Laplacian to perform neighborhood aggregation. It computes updated node features by linearly transforming input features and weighting them based on the normalized adjacency matrix, effectively averaging over immediate neighbors \cite{kipf_semi-supervised_2017}. The GCN model uses three GCNConv layers and incorporates dropout regularization. The GATConv layer extends this mechanism through masked self-attention, computing attention coefficients for each edge based on the concatenated feature vectors of connected nodes. In our implementation, we use multi-head attention with four heads in the first two layers and a single output head in the final layer. This allows the model to jointly attend to information from different representation subspaces. To ensure a controlled comparison, all models were initially tested using ReLU (Rectified Linear Unit) activations. However, given empirical evidence favoring ELU (Exponential Linear Unit) in attention-based networks, we also evaluated GAT with ELU, as proposed in the original architecture \cite{velickovic_graph_2018}. The SAGEConv layer, from the GraphSAGE framework, aggregates features from sampled neighborhood nodes using a mean aggregator. It concatenates the target node’s own features with the aggregated neighbor features before applying a learnable transformation and non-linearity \cite{hamilton_inductive_2017}. All layers are followed by non-linear activation functions, primarily ReLU or ELU, and optionally interleaved with dropout for regularization. These design choices enable a controlled comparison of how different message-passing strategies affect classification performance in both content-based and content-agnostic input settings.\\

\subsection*{Hyperparameter Tuning}

\begin{table}[!ht]
\centering
\caption{Hyperparameter Grid for Grid Search}
\label{tab:param_grid}
\begin{tabular}{ll}
\toprule
\textbf{Hyperparameter} & \textbf{Values} \\
\midrule
Hidden Size   & 32, 64, 128 \\
Dropout Rate  & 0.0, 0.3, 0.5 \\
Learning Rate & 0.01, 0.005, 0.001 \\
Weight Decay  & $1 \times 10^{-5}$, $1 \times 10^{-4}$, $1 \times 10^{-3}$ \\
\bottomrule
\end{tabular}
\end{table}

This grid results in a total of $3\times3\times3\times3=81$ distinct configurations. The GNN models as well as the MLP were trained with 100 epochs and evaluated independently using a fixed training and validation split to ensure comparability across experiments. To ensure a consistent comparison across all models, early stopping was not used and training was conducted for a fixed number of 100 epochs. In all cases we observe convergence of the models and no sign of overfitting, more details are reported in the SI. The table \ref{tab:best_params} summarizes the best hyperparameter configurations and corresponding validation losses for all models under both content and content-agnostic settings.

\begin{table}[h!]
\centering
\caption{Best Hyperparameters for All Models }
\label{tab:best_params}
\begin{tabular}{l c c c c c}
\toprule
Model & dropout & hidden size& learning rate& weight decay& val loss\\
\midrule
\multicolumn{6}{c}{\textbf{Content}} \\
\midrule
GCN               & 0 & 64 & 0.001 & 0.001  & 0.85 \\
GAT               & 0 & 128 & 0.01  & 0.001  & 0.88 \\
GraphSAGE         & 0.3 & 64 & 0.001 & 0.001  & 0.83 \\
MLP               & 0.5 & 64 & 0.005 & 0.001  & 0.85 \\
\midrule
\multicolumn{6}{c}{\textbf{Content-Agnostic}} \\
\midrule
GCN               & 0.3 & 128 & 0.01  & 1e-05  & 0.91 \\
GAT               & 0 & 128 & 0.001 & 1e-05  & 0.92 \\
GraphSAGE         & 0 & 128 & 0.01  & 1e-05  & 0.93 \\
MLP               & 0 & 32 & 0.01  & 0.0001  & 0.97 \\
\bottomrule
\end{tabular}
\end{table}

\section*{Supplementary information}
Additional analysis and robustness tests are reported in the Supplementary Information

\section*{Declarations}
\subsection*{Funding}
The authors have no funding to declare.

\subsection*{Code availability}
All code is publicly available at \url{https://github.com/Raphaaella/GNN_Telegram}.

\subsection*{Acknowledgments}
We are grateful to Prof. David Garcia for interesting discussions. 


\clearpage
\appendix
\setcounter{figure}{0}
\setcounter{table}{0}
\setcounter{section}{0}
\renewcommand{\thefigure}{S\arabic{figure}}
\renewcommand{\thetable}{S\arabic{table}}
\renewcommand{\thesection}{S\arabic{section}}

\begin{center}
{\Large\bfseries Supplementary Information}
\end{center}
\vspace{1em}

\section{Dataset Description and Descriptive Statistics}
\label{sec:si_dataset}

\subsection{Most Frequently Shared News Domains}

\begin{figure}[H]
    \centering
    \includegraphics[width=0.75\linewidth]{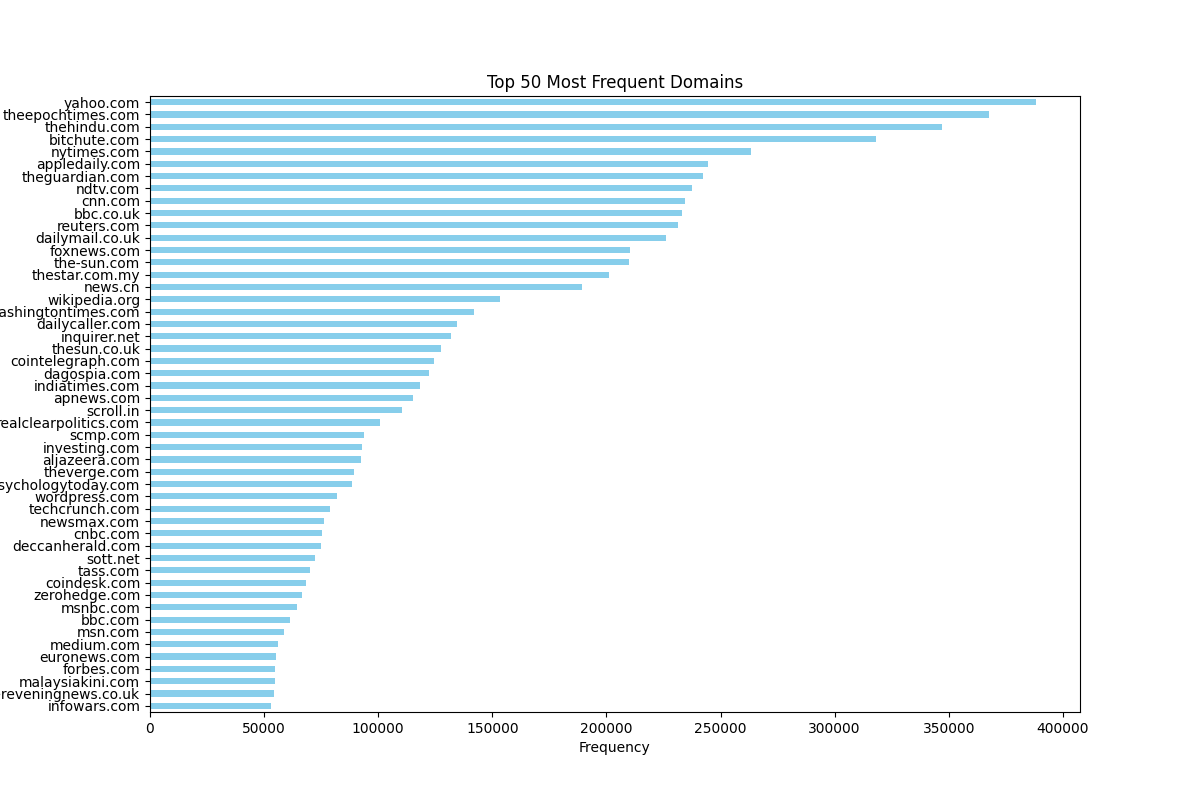}
    \caption{Top 50 most frequently shared domains in the Telegram dataset, ranked by the total number of URL occurrences in the dataset (computed before sampling at most 1000 URLs per domain).} 
    \label{fig:si_top50}
\end{figure}

To provide an overview of the news ecosystem captured by our Telegram sample, Fig.~\ref{fig:si_top50} shows the 50 most frequently shared news domains together with their reliability classification, computed before the sampling of at most 1000 URLs per domain (see Methods in the main text). The list is dominated by a mixture of mainstream outlets (e.g., \texttt{nytimes.com}, \texttt{theguardian.com}, \texttt{reuters.com}) and well-known low-reliability sources. Notably, \texttt{epochtimes.com} is among the most shared news domains in the dataset despite having a reliability score below $0.33$, i.e., falling into the ``Unreliable'' category. This illustrates, at the level of individual sources, the high prevalence of low-reliability information on Telegram discussed in the main text.

\subsection{Distribution of URLs per Domain}

\begin{figure}[H]
    \centering
    \includegraphics[width=0.7\linewidth]{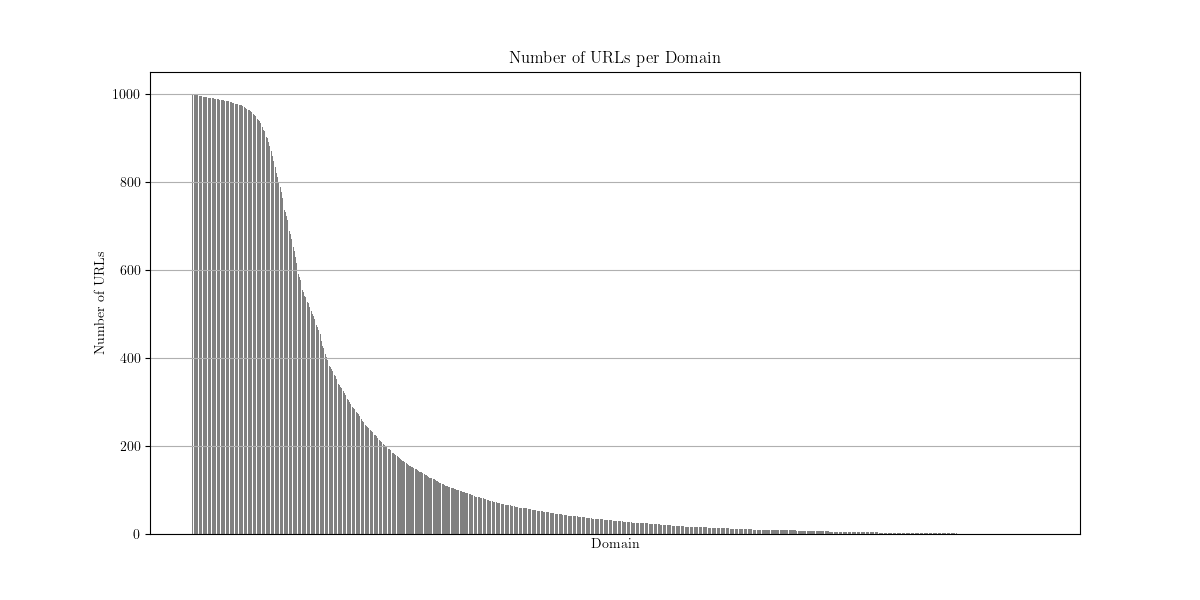}
    \caption{Number of successfully scraped URLs per domain, after sampling at most 1000 URLs per domain. Domains are sorted in decreasing order.}
    \label{fig:si_urls_per_domain}
\end{figure}

For the content-based features, we retained the 1000 most frequently forwarded URLs per domain and scraped the corresponding articles. Fig.~\ref{fig:si_urls_per_domain} shows the resulting distribution of successfully scraped URLs per domain. The distribution is heavily right-skewed: while a minority of highly active domains reach the 1000-URL cap, most domains are represented by substantially fewer articles. Part of this incompleteness stems from URLs that had expired by the time of scraping. Since domain-level content embeddings are obtained by averaging over all available article embeddings, domains with few scraped articles have noisier content representations; the network structure and the spreading features are unaffected by this limitation.

\subsection{Descriptive Statistics of the Spreading Features}

Table~\ref{tab:si_context_stats} reports descriptive statistics of the domain-level spreading (content-agnostic) features prior to z-score standardization. The distributions are highly heterogeneous: while the median domain is shared in fewer than two chats and generates only a couple of messages, the most active domains reach thousands of messages and hundreds of avalanches. The virality measure is concentrated at high values, indicating that for most domains the temporal spreading pattern is dominated by a single large avalanche.

\begin{table}[H]
\centering
\caption{Descriptive statistics of the domain-level spreading features (training set, $N = 4884$ domains, before z-score normalization).}
\label{tab:si_context_stats}
\begin{tabular}{lcccc}
\toprule
 & \textbf{Virality (\%)} & \textbf{Avalanches} & \textbf{Messages} & \textbf{Chats} \\
\midrule
Mean            & 85.00  & 4.70    & 10.21   & 2.78 \\
Std.\ Dev.      & 14.00  & 27.91   & 107.47  & 7.25 \\
Min             & 3.00   & 1.00    & 1.00    & 1.00 \\
25th Percentile & 81.00  & 1.33    & 1.50    & 1.16 \\
Median          & 88.00  & 2.00    & 2.45    & 1.57 \\
75th Percentile & 94.00  & 3.18    & 4.11    & 2.26 \\
Max             & 100.00 & 1560.00 & 5436.62 & 196.67 \\
\bottomrule
\end{tabular}
\end{table}

\begin{figure}[H]
    \centering
    \includegraphics[width=0.7\linewidth]{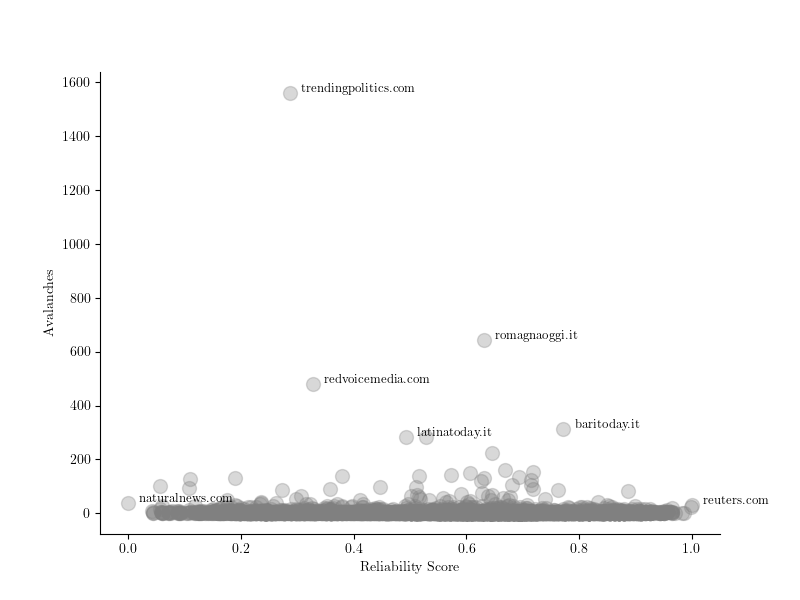}
    \caption{Number of avalanches per domain versus the domain reliability score (PC1). Selected outlier domains are annotated.}
    \label{fig:si_avalanches}
\end{figure}

Fig.~\ref{fig:si_avalanches} relates one of these features, the number of avalanches, to the domain reliability score. Most domains cluster at low avalanche counts across the whole reliability range. However, a small number of outlier domains, such as \texttt{trendingpolitics.com} and \texttt{redvoicemedia.com}, exhibit extremely high avalanche frequencies, consistent with aggressive or campaign-like dissemination strategies. The presence of such outliers motivates the inclusion of avalanche-based features among the content-agnostic node attributes.

\section{Content Embeddings}
\label{sec:si_embeddings}

\begin{figure}[H]
    \centering
    \includegraphics[width=0.7\linewidth]{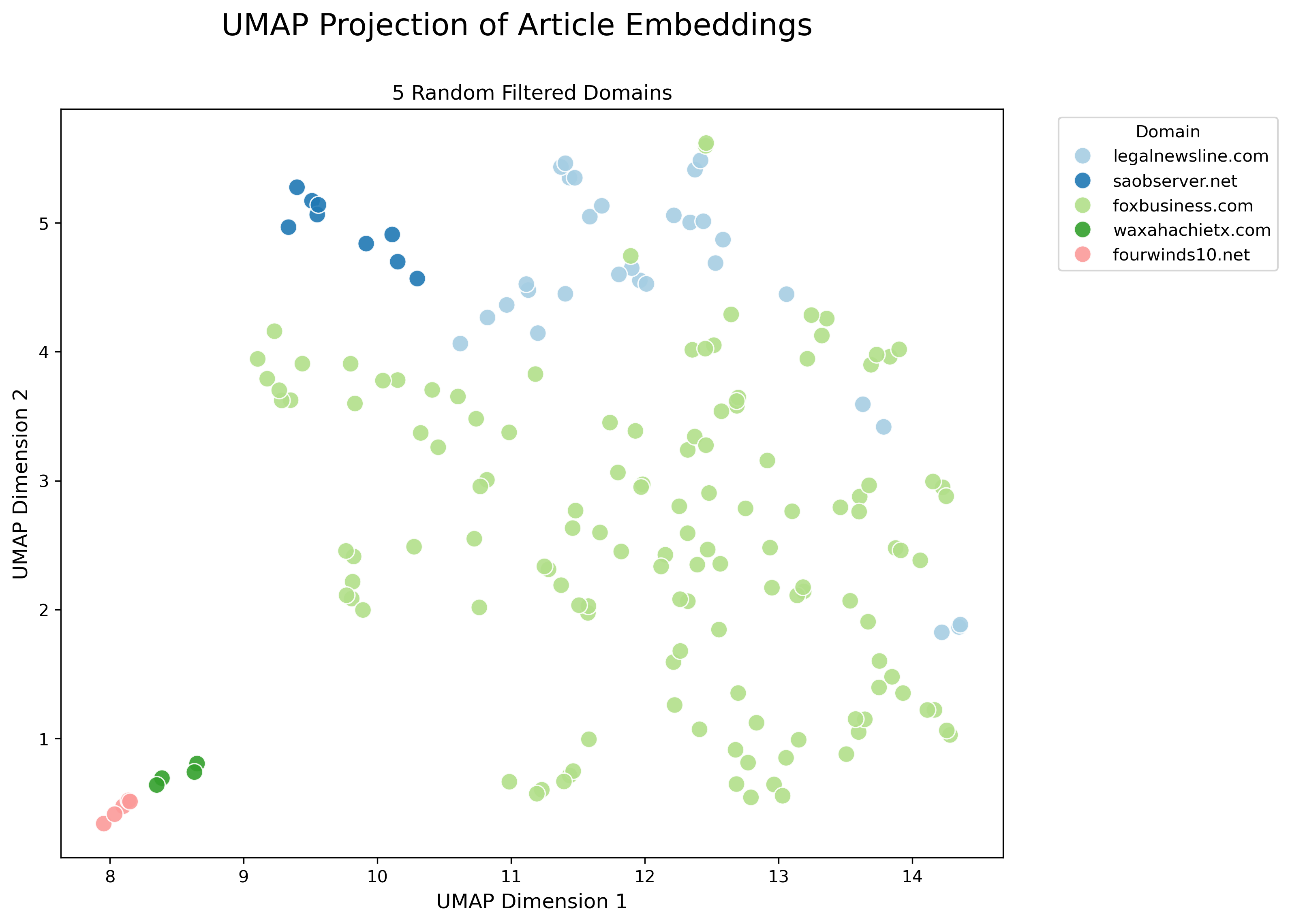}
    \caption{UMAP projection of the SBERT article embeddings (paraphrase-multilingual-MiniLM-L12-v2, 384 dimensions) for ten randomly selected domains. Colors indicate the domain of origin.}
    \label{fig:si_umap}
\end{figure}

To assess whether domain-averaged text embeddings constitute a meaningful content representation, Fig.~\ref{fig:si_umap} shows a two-dimensional UMAP projection \cite{mcinnes_umap_2020} of the article-level SBERT embeddings for ten randomly selected domains. Articles from the same domain form dense, well-separated clusters, reflecting a strong within-domain semantic coherence in topics and style. This observation supports our choice of representing each domain by the average embedding of its articles: the domain-level centroid is a good summary of the underlying article distribution.

\section{Robustness of the Network Validation}
\label{sec:si_bicm}

The statistically validated domain--domain network used in the main text was obtained by pruning the monopartite projection with the Bipartite Configuration Model at a significance threshold of $p = 0.01$. To verify that our results do not depend on this specific choice, we repeated the validation procedure with alternative thresholds ($p = 0.05$ and $p = 0.5$), which yield progressively denser networks. Training the GNN models on these alternative networks did not improve classification performance, indicating that the results reported in the main text are robust with respect to the pruning threshold and that the $p = 0.01$ backbone already retains the structurally informative connections.

\section{Additional Model Variants and Baselines}
\label{sec:si_models}

\subsection{GCN Depth and Regularization}

In addition to the three GNN architectures reported in the main text, we evaluated two simpler GCN variants in order to isolate the effect of depth and dropout regularization: a two-layer GCN without dropout (``GCN-2''), and a three-layer GCN without dropout (``GCN-3''). Table~\ref{tab:si_architectures} summarizes all evaluated architectures, and Tables~\ref{tab:si_results_content} and \ref{tab:si_results_agnostic} report their performance. Interestingly, in the content-based setting the shallow two-layer GCN slightly outperforms both three-layer GCN variants, suggesting that additional message-passing depth provides no benefit on this network and may instead promote over-smoothing or overfitting. GraphSAGE remains the best-performing architecture in all settings.

\begin{table}[H]
\centering
\caption{Summary of all evaluated model architectures.}
\label{tab:si_architectures}
\begin{tabular}{lcllc}
\toprule
\textbf{Model} & \textbf{Layer Count} & \textbf{Layer Type} & \textbf{Activation} & \textbf{Dropout} \\
\midrule
GCN-2 (super basic) & 2 & GCNConv  & ReLU & No  \\
GCN-3 (basic)       & 3 & GCNConv  & ReLU & No  \\
GCN                 & 3 & GCNConv  & ReLU & Yes \\
GAT                 & 3 & GATConv  & ELU  & Yes \\
GraphSAGE           & 3 & SAGEConv & ReLU & Yes \\
MLP                 & 3 & Linear   & ReLU & Yes \\
SBERT classifier    & -- & Transformer + softmax head & -- & Yes \\
\bottomrule
\end{tabular}
\end{table}

\subsection{Transformer-Based Content Baseline}

As an additional network-unaware baseline for the content-based setting, we trained a purely text-based classifier on top of the same SBERT model used for feature extraction (paraphrase-multilingual-MiniLM-L12-v2). For each domain, the scraped articles were concatenated, lower-cased, stripped of special characters and digits, tokenized, and truncated to 512 tokens; the classifier consists of a dropout layer followed by a fully connected softmax head, optimized with Adam and negative log-likelihood loss. Hyperparameters (learning rate, batch size, weight decay, warmup steps) were tuned by grid search.

This transformer baseline reaches an accuracy of 0.49 and an F1-score of 0.49 (Table~\ref{tab:si_results_content}), which is below the MLP baseline operating on the combined feature set and roughly 14 percentage points below GraphSAGE. Its confusion matrix (Fig.~\ref{fig:si_content_curves_2}) shows that it detects only 39 of the 171 unreliable domains in the test set, compared to 101 for GraphSAGE. Two factors likely contribute to this gap: (i) the truncation to 512 tokens discards most of the concatenated per-domain text, and (ii) the classifier has no access to the co-sharing network or spreading dynamics. This result reinforces the conclusion of the main text: content alone, even when processed by a dedicated transformer classifier, is insufficient, and network structure carries substantial complementary signal.

\subsection{Full Results}

\begin{table}[H]
\centering
\caption{Performance of all models in the content-based setting (spreading features + content embeddings). Best values in bold.}
\label{tab:si_results_content}
\begin{tabular}{lcccc}
\toprule
\textbf{Model} & \textbf{Accuracy} & \textbf{Precision} & \textbf{Recall} & \textbf{F1-Score} \\
\midrule
GCN-2 (super basic) & 0.5851 & 0.5832 & 0.5851 & 0.5830 \\
GCN-3 (basic)       & 0.5843 & 0.5846 & 0.5843 & 0.5840 \\
GCN                 & 0.5728 & 0.5696 & 0.5728 & 0.5706 \\
GAT                 & 0.5696 & 0.5773 & 0.5696 & 0.5637 \\
GraphSAGE           & \textbf{0.6293} & \textbf{0.6285} & \textbf{0.6293} & \textbf{0.6262} \\
MLP Baseline        & 0.5499 & 0.5600 & 0.5499 & 0.5518 \\
SBERT Baseline      & 0.4900 & 0.4800 & 0.4810 & 0.4900 \\
\bottomrule
\end{tabular}
\end{table}

\begin{table}[H]
\centering
\caption{Performance of all models in the content-agnostic setting (spreading features only). Best values in bold.}
\label{tab:si_results_agnostic}
\begin{tabular}{lcccc}
\toprule
\textbf{Model} & \textbf{Accuracy} & \textbf{Precision} & \textbf{Recall} & \textbf{F1-Score} \\
\midrule
GCN-2 (super basic) & 0.4877 & 0.4907 & 0.4877 & 0.4683 \\
GCN-3 (basic)       & 0.4853 & 0.5220 & 0.4853 & 0.4550 \\
GCN                 & 0.4967 & 0.5283 & 0.4967 & 0.4727 \\
GAT                 & 0.4992 & 0.5174 & 0.4992 & 0.4766 \\
GraphSAGE           & \textbf{0.5311} & \textbf{0.5367} & \textbf{0.5311} & \textbf{0.5196} \\
MLP Baseline        & 0.4697 & 0.4267 & 0.4697 & 0.4300 \\
\bottomrule
\end{tabular}
\end{table}

While GraphSAGE achieves the best overall accuracy also in the content-agnostic setting, we note that the three-layer GCN without dropout attains the highest number of correctly identified unreliable domains in this setting (46 out of 171, versus 42 for GraphSAGE), at the cost of lower overall accuracy. Depending on the application, e.g.\ when minimizing false negatives on the unreliable class is the primary objective, such recall-oriented model choices may be preferable.

\section{Learning Curves and Confusion Matrices}
\label{sec:si_curves}

For completeness, this section reports the training and validation loss curves as well as the test-set confusion matrices for all models in both feature settings. All models were trained for a fixed 100 epochs without early stopping to ensure comparability; the loss curves show convergence and no signs of severe overfitting. Note the confusion matrix of the content-agnostic MLP (Fig.~\ref{fig:si_agnostic_curves_2}): the model never predicts the unreliable class, i.e., it produces false negatives for all 171 unreliable test domains. In contrast, all graph-aware models recover a substantial fraction of this class, illustrating concretely the practical value of network information for misinformation detection.

\subsection{Content-Based Setting}

\begin{figure}[H]
    \centering
    \subfigure[GCN-2 - Train/Val]{\includegraphics[width=0.40\linewidth]{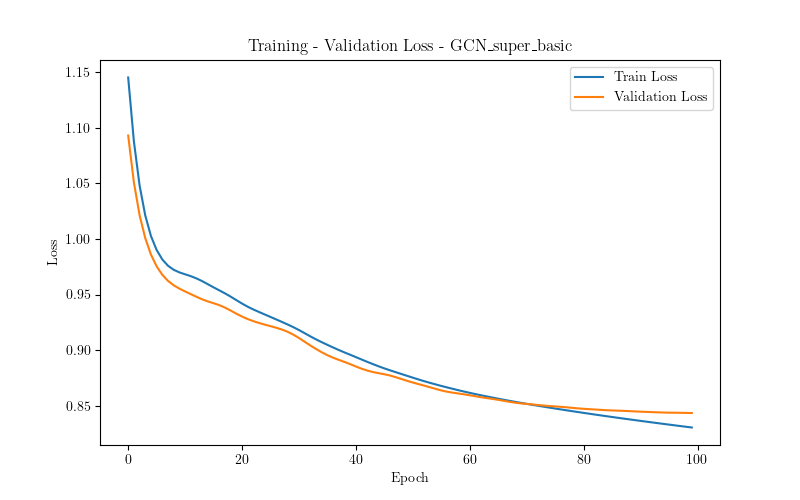}}
    \subfigure[GCN-2 - Confusion]{\includegraphics[width=0.40\linewidth]{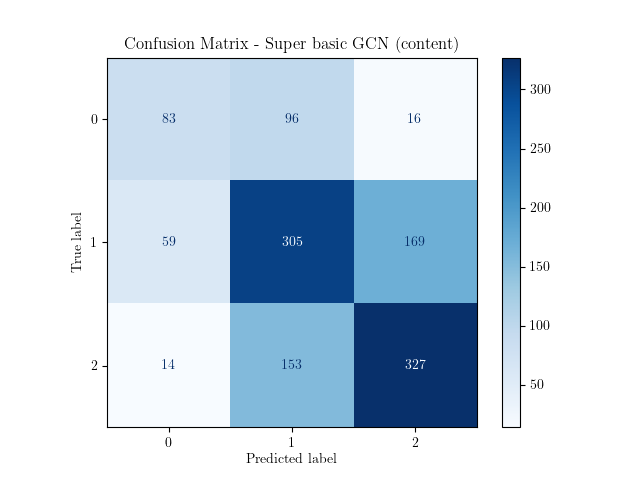}}

    \subfigure[GCN-3 - Train/Val]{\includegraphics[width=0.40\linewidth]{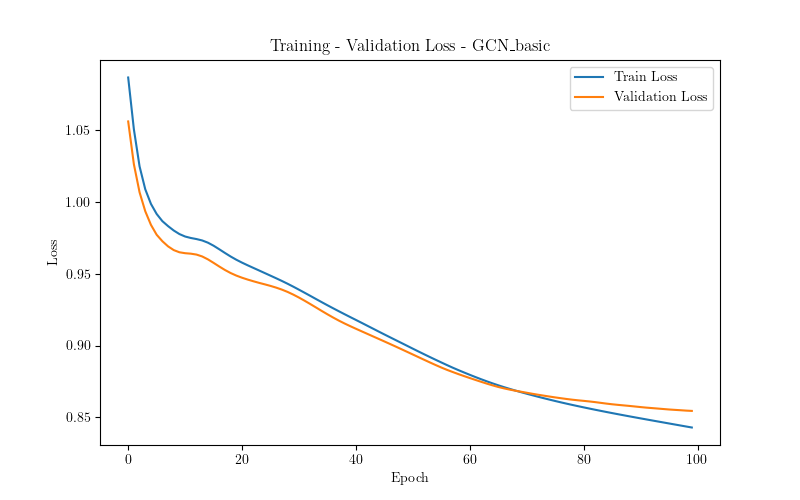}}
    \subfigure[GCN-3 - Confusion]{\includegraphics[width=0.40\linewidth]{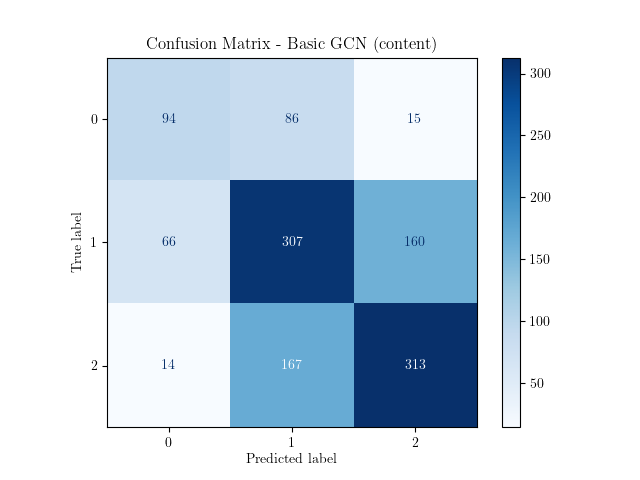}}

    \subfigure[GCN - Train/Val]{\includegraphics[width=0.40\linewidth]{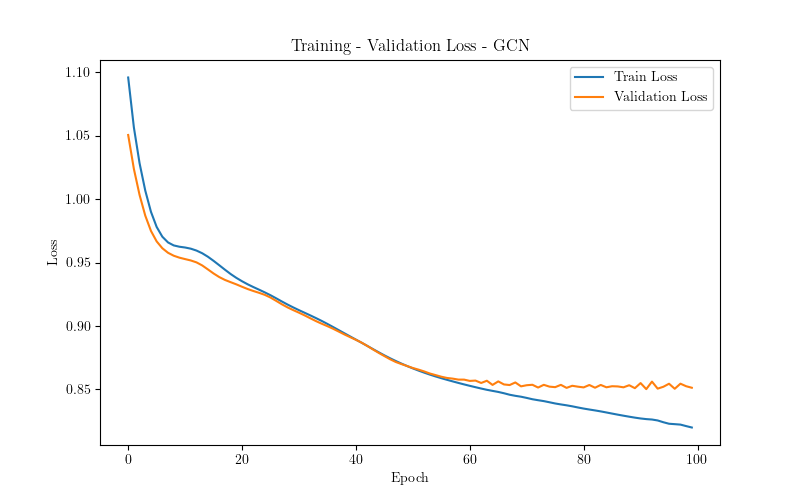}}
    \subfigure[GCN - Confusion]{\includegraphics[width=0.40\linewidth]{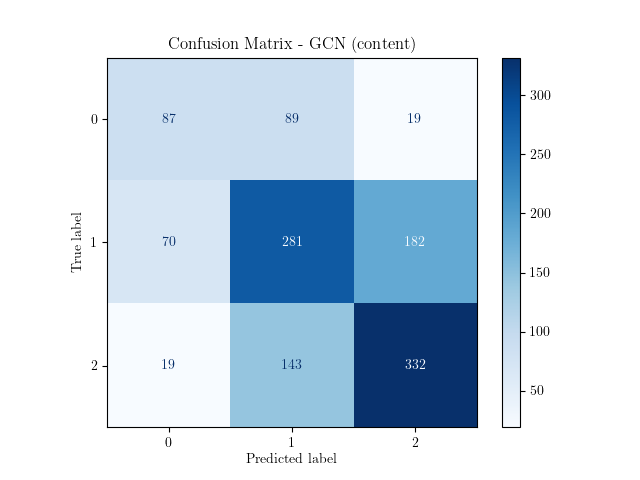}}

    \caption{Train-validation loss curves and confusion matrices for the GCN variants (content-based setting).}
    \label{fig:si_content_curves_1}
\end{figure}

\begin{figure}[H]
    \centering
    \subfigure[GAT - Train/Val]{\includegraphics[width=0.40\linewidth]{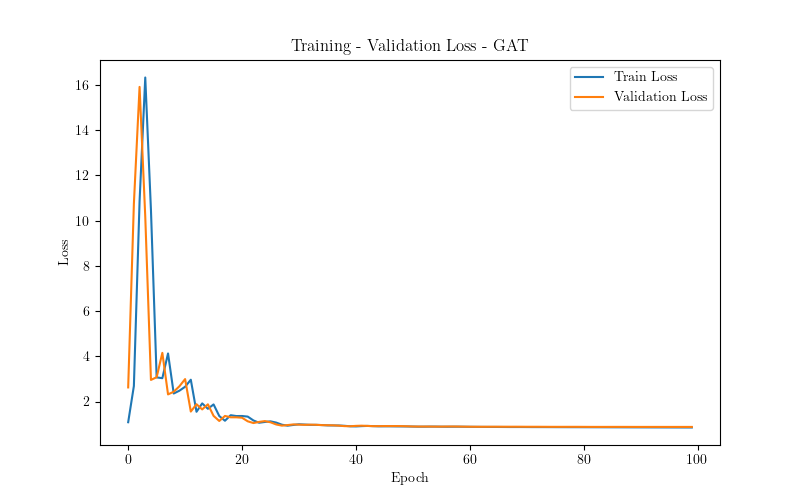}}
    \subfigure[GAT - Confusion]{\includegraphics[width=0.40\linewidth]{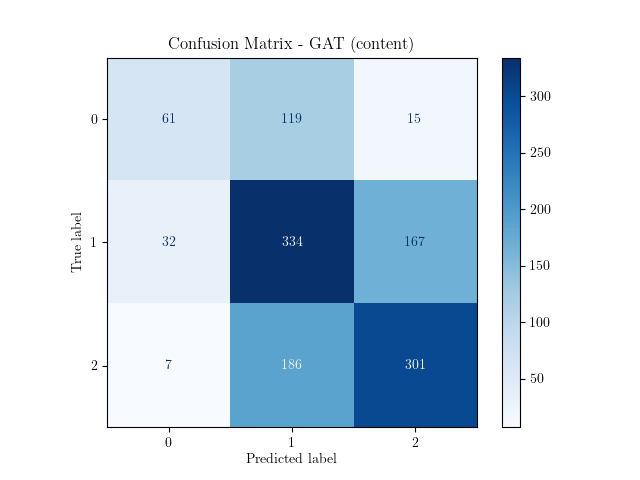}}

    \subfigure[GraphSAGE - Train/Val]{\includegraphics[width=0.40\linewidth]{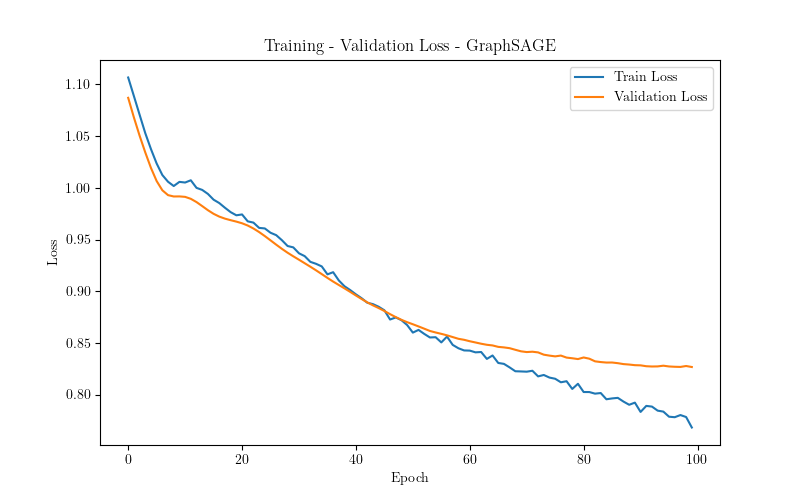}}
    \subfigure[GraphSAGE - Confusion]{\includegraphics[width=0.40\linewidth]{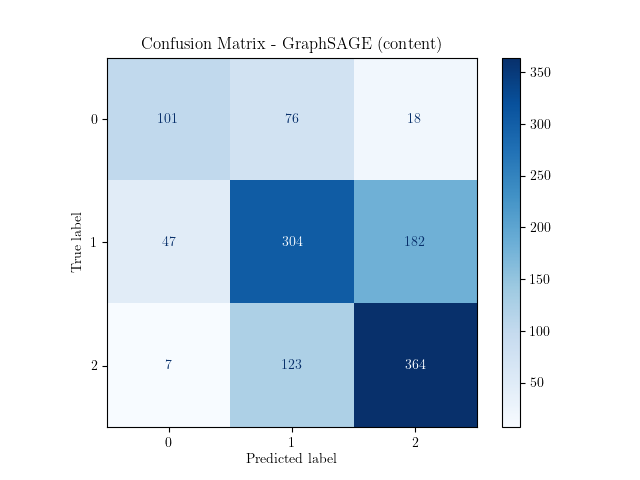}}

    \subfigure[MLP - Train/Val]{\includegraphics[width=0.40\linewidth]{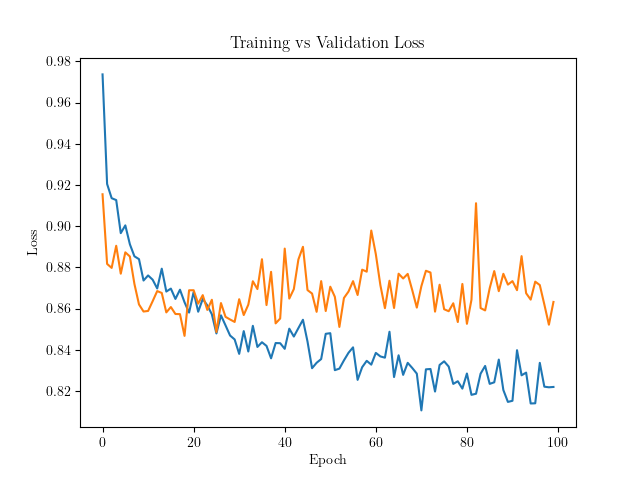}}
    \subfigure[MLP - Confusion]{\includegraphics[width=0.40\linewidth]{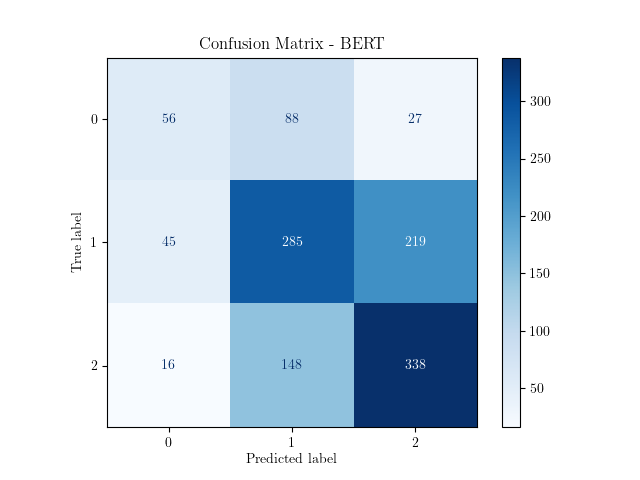}}

    \subfigure[SBERT - Train/Val]{\includegraphics[width=0.40\linewidth]{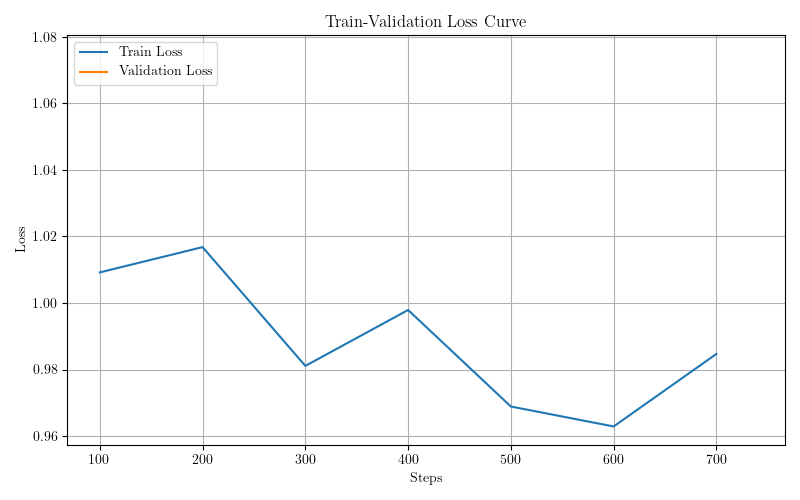}}
    \subfigure[SBERT - Confusion]{\includegraphics[width=0.40\linewidth]{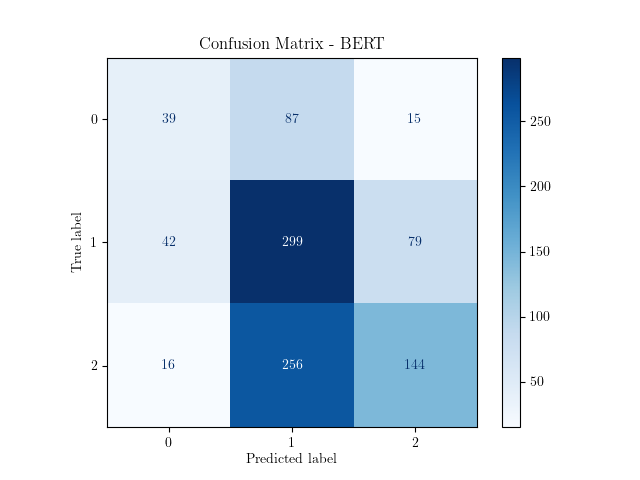}}

    \caption{Train-validation loss curves and confusion matrices for GAT, GraphSAGE, MLP, and the SBERT baseline (content-based setting).}
    \label{fig:si_content_curves_2}
\end{figure}

\subsection{Content-Agnostic Setting}

\begin{figure}[H]
    \centering
    \subfigure[GCN-2 - Train/Val]{\includegraphics[width=0.40\linewidth]{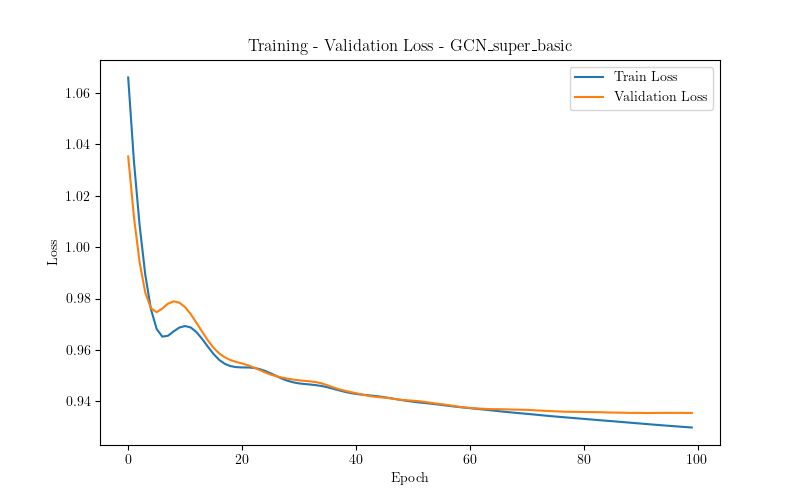}}
    \subfigure[GCN-2 - Confusion]{\includegraphics[width=0.40\linewidth]{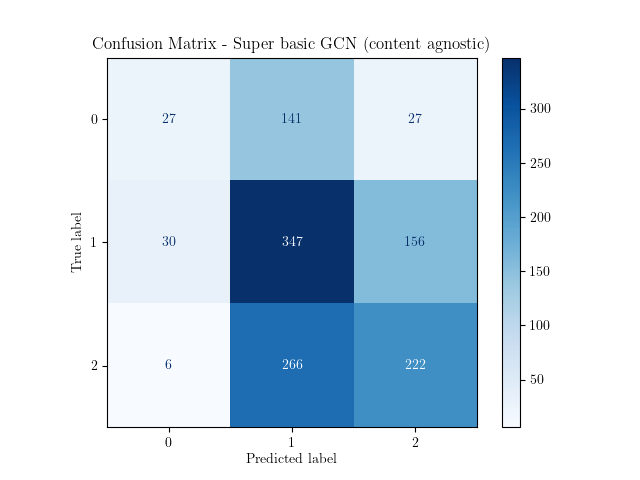}}

    \subfigure[GCN-3 - Train/Val]{\includegraphics[width=0.40\linewidth]{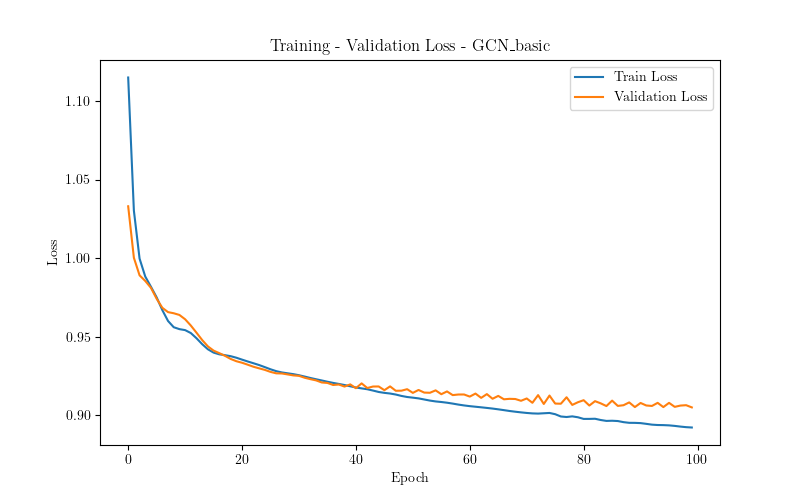}}
    \subfigure[GCN-3 - Confusion]{\includegraphics[width=0.40\linewidth]{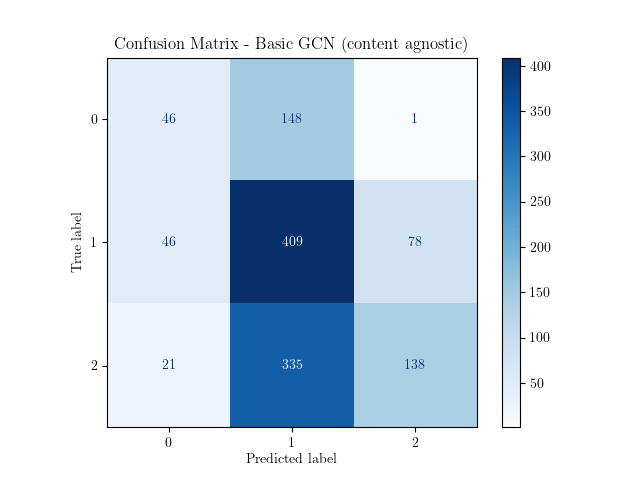}}

    \subfigure[GCN - Train/Val]{\includegraphics[width=0.40\linewidth]{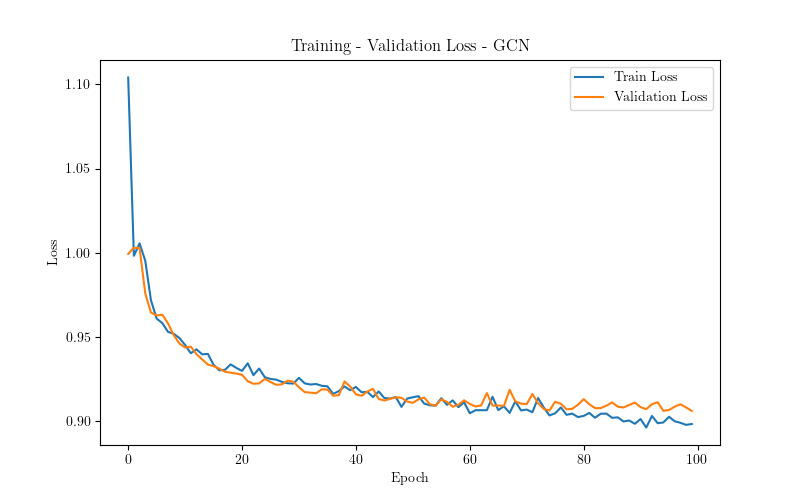}}
    \subfigure[GCN - Confusion]{\includegraphics[width=0.40\linewidth]{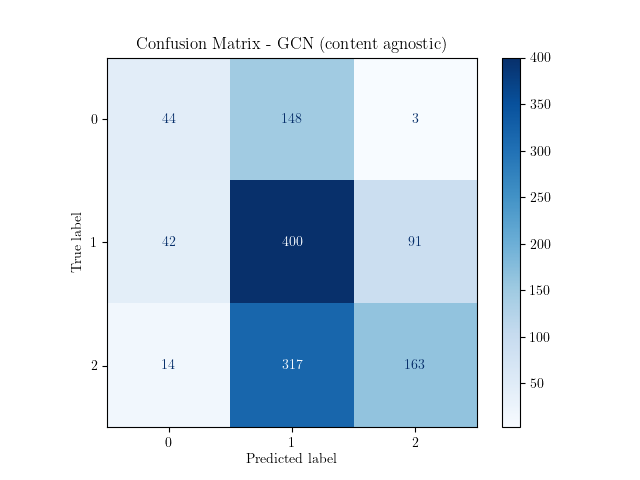}}

    \caption{Train-validation loss curves and confusion matrices for the GCN variants (content-agnostic setting).}
    \label{fig:si_agnostic_curves_1}
\end{figure}

\begin{figure}[H]
    \centering
    \subfigure[GAT - Train/Val]{\includegraphics[width=0.40\linewidth]{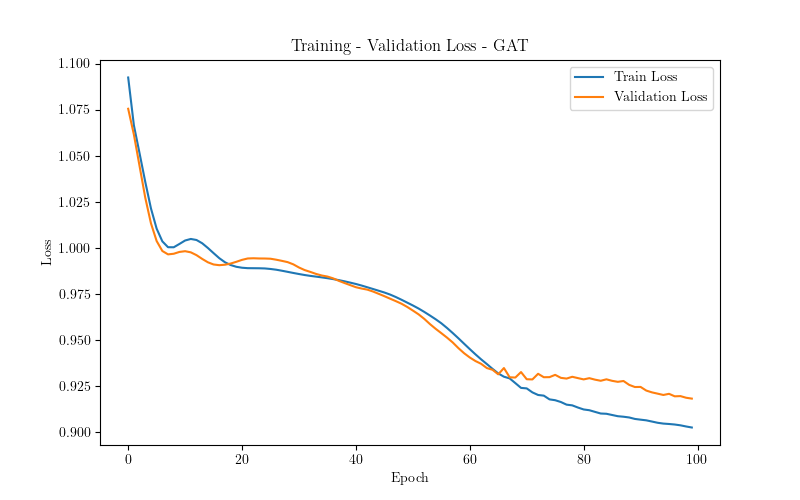}}
    \subfigure[GAT - Confusion]{\includegraphics[width=0.40\linewidth]{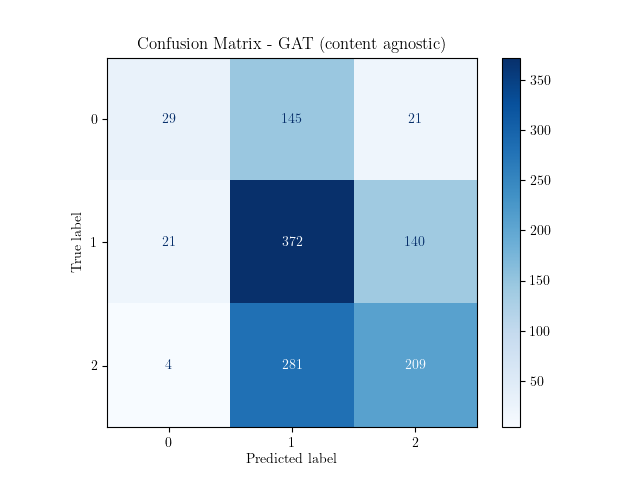}}

    \subfigure[GraphSAGE - Train/Val]{\includegraphics[width=0.40\linewidth]{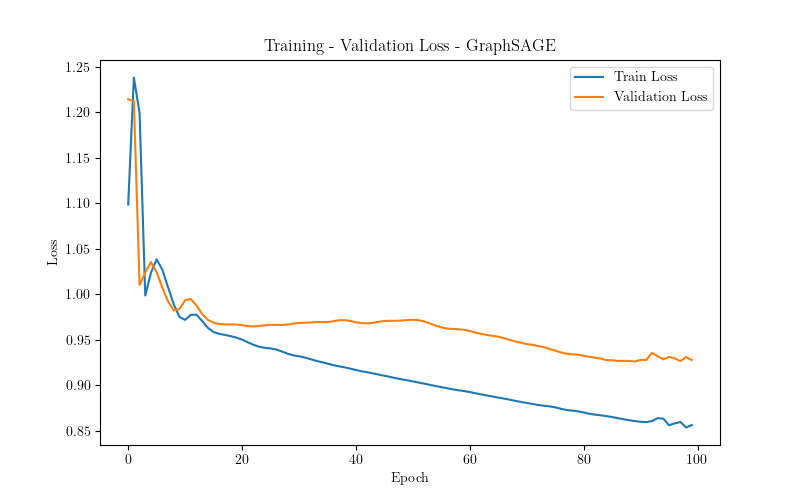}}
    \subfigure[GraphSAGE - Confusion]{\includegraphics[width=0.40\linewidth]{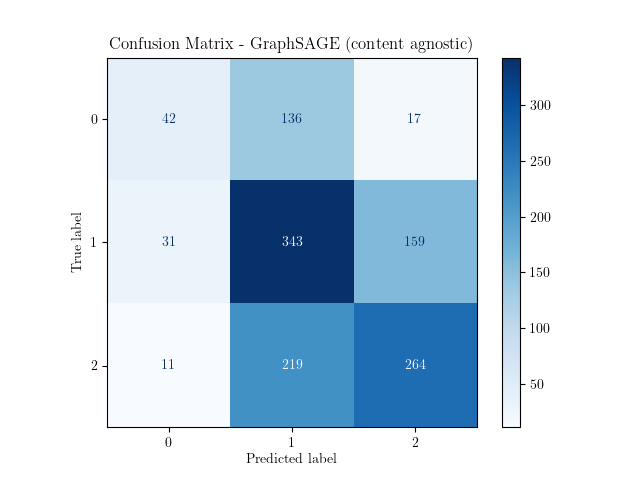}}

    \subfigure[MLP - Train/Val]{\includegraphics[width=0.40\linewidth]{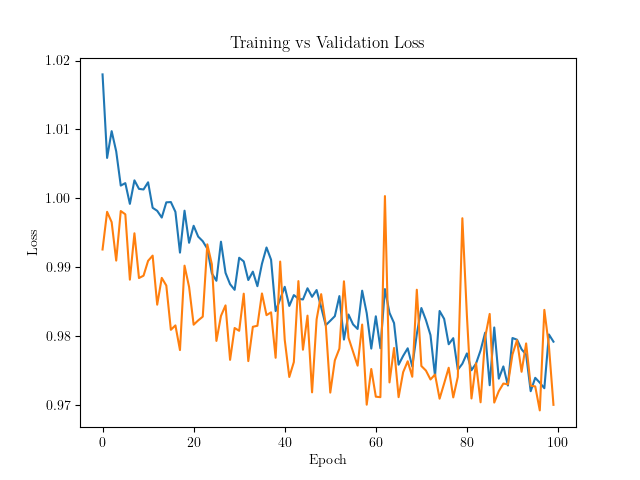}}
    \subfigure[MLP - Confusion]{\includegraphics[width=0.40\linewidth]{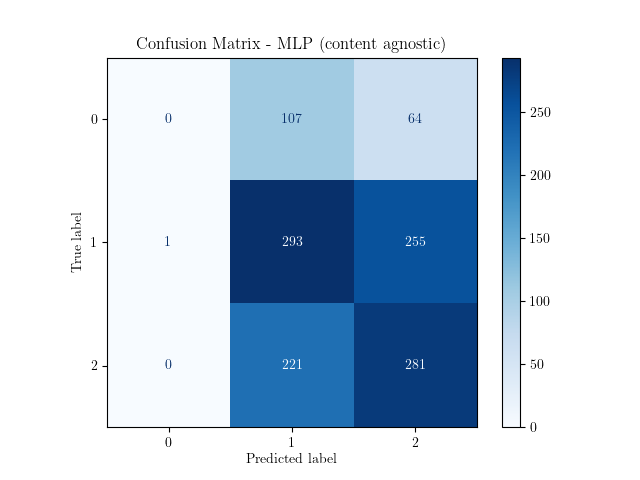}}

    \caption{Train-validation loss curves and confusion matrices for GAT, GraphSAGE, and MLP (content-agnostic setting). Note that the MLP never predicts the unreliable class.}
    \label{fig:si_agnostic_curves_2}
\end{figure}


\begin{thebibliography}{10}

\bibitem{carragher_detection_2024}
Peter Carragher, Evan~M. Williams, and Kathleen~M. Carley.
\newblock Detection and {Discovery} of {Misinformation} {Sources} {Using}
  {Attributed} {Webgraphs}.
\newblock {\em Proceedings of the International AAAI Conference on Web and
  Social Media}, 18:214--226, May 2024.

\bibitem{ceci_messaging_2025}
Laura Ceci.
\newblock Messaging apps: most popular by global downloads 2025, February 2025.

\bibitem{cinelli_echo_2021}
Matteo Cinelli, Gianmarco De~Francisci~Morales, Alessandro Galeazzi, Walter
  Quattrociocchi, and Michele Starnini.
\newblock The echo chamber effect on social media.
\newblock {\em Proceedings of the National Academy of Sciences},
  118(9):e2023301118, March 2021.

\bibitem{cinelli_covid-19_2020}
Matteo Cinelli, Walter Quattrociocchi, Alessandro Galeazzi, Carlo~Michele
  Valensise, Emanuele Brugnoli, Ana~Lucia Schmidt, Paola Zola, Fabiana Zollo,
  and Antonio Scala.
\newblock The {COVID}-19 social media infodemic.
\newblock {\em Scientific Reports}, 10(1):16598, October 2020.

\bibitem{collins-dictionary_collins_2017}
Collins-Dictionary.
\newblock Collins 2017 {Word} of the {Year} {Shortlist}, November 2017.

\bibitem{langdetectport}
Michal Danilák.
\newblock langdetect: Port of google's language-detection library to python.
\newblock \url{https://github.com/Mimino666/langdetect}, 2021.

\bibitem{durov_du_2018}
Pavel Durov.
\newblock Du {Rove}'s {Channel} - {Digital} {Resistance}, 2018.

\bibitem{durov_du_2025}
Pavel Durov.
\newblock Du {Rove}'s {Channel} - {Telegram} monthly active user, March 2025.

\bibitem{fey2019fast}
Matthias Fey and Jan~Eric Lenssen.
\newblock Fast graph representation learning with pytorch geometric.
\newblock {\em arXiv preprint arXiv:1903.02428}, 2019.

\bibitem{goertzel_belief_1994}
Ted Goertzel.
\newblock Belief in {Conspiracy} {Theories}.
\newblock {\em Political Psychology}, 15(4):731--742, 1994.
\newblock Publisher: [International Society of Political Psychology, Wiley].

\bibitem{TeraGram}
Anastasia Golovin, Sebastian~B. Mohr, Arne~I. Gottwald, Ulrik Hvid, Srushhti
  Trivedi, Joao Pinheiro~Neto, Andreas~C. Schneider, and Viola Priesemann.
\newblock Teragram: A structured longitudinal dataset of the telegram
  messenger.
\newblock {\em Proceedings of the International AAAI Conference on Web and
  Social Media}, 20(1):2794–2816, May 2026.

\bibitem{gul_advancing_2024}
Haji Gul, Feras Al-Obeidat, Muhammad Wasim, Adnan Amin, and Fernando Moreira.
\newblock Advancing {Fake} {News} {Detection} with {Graph} {Neural} {Network}
  and {Deep} {Learning}.
\newblock {\em Journal of Physics: Complexity}, August 2024.

\bibitem{hamilton_inductive_2017}
Will Hamilton, Zhitao Ying, and Jure Leskovec.
\newblock Inductive {Representation} {Learning} on {Large} {Graphs}.
\newblock {\em Advances in Neural Information Processing Systems}, 30, 2017.

\bibitem{han_graph_2020}
Yi~Han, Shanika Karunasekera, and Christopher Leckie.
\newblock Graph {Neural} {Networks} with {Continual} {Learning} for {Fake}
  {News} {Detection} from {Social} {Media}, August 2020.
\newblock arXiv:2007.03316 [cs].

\bibitem{herasimenka_misinformation_2023}
Aliaksandr Herasimenka, Jonathan Bright, Aleksi Knuutila, and Philip~N. Howard.
\newblock Misinformation and professional news on largely unmoderated
  platforms: the case of telegram.
\newblock {\em Journal of Information Technology \& Politics}, 20(2):198--212,
  April 2023.

\bibitem{hornik_association_2021}
Robert Hornik, Kikut ~, Ava, Jesch ~, Emma, Woko ~, Chioma, Siegel ~, Leeann, ,
  and Kwanho Kim.
\newblock Association of {COVID}-19 {Misinformation} with {Face} {Mask}
  {Wearing} and {Social} {Distancing} in a {Nationally} {Representative} {US}
  {Sample}.
\newblock {\em Health Communication}, 36(1):6--14, January 2021.
\newblock Publisher: Routledge \_eprint:
  https://doi.org/10.1080/10410236.2020.1847437.

\bibitem{hugging-face_sentence-transformersparaphrase-multilingual-minilm-l12-v2_2019}
Hugging-Face.
\newblock sentence-transformers/paraphrase-multilingual-{MiniLM}-{L12}-v2 ·
  {Hugging} {Face}, 2019.

\bibitem{khan_fake_2021}
Tanveer Khan, Antonis Michalas, and Adnan Akhunzada.
\newblock Fake news outbreak 2021: {Can} we stop the viral spread?
\newblock {\em Journal of Network and Computer Applications}, 190:103112,
  September 2021.

\bibitem{kipf_semi-supervised_2017}
Thomas~N. Kipf and Max Welling.
\newblock Semi-{Supervised} {Classification} with {Graph} {Convolutional}
  {Networks}, February 2017.
\newblock arXiv:1609.02907.

\bibitem{ksiazek_user_2016}
Thomas~B Ksiazek, Limor Peer, and Kevin Lessard.
\newblock User engagement with online news: {Conceptualizing} interactivity and
  exploring the relationship between online news videos and user comments.
\newblock {\em New Media \& Society}, 18(3):502--520, March 2016.

\bibitem{lin_high_2023}
Hause Lin, Jana Lasser, Stephan Lewandowsky, Rocky Cole, Andrew Gully, David~G
  Rand, and Gordon Pennycook.
\newblock High level of correspondence across different news domain quality
  rating sets.
\newblock {\em PNAS Nexus}, 2(9):pgad286, September 2023.

\bibitem{Luhring2025}
Jula L\"{u}hring, Hannah Metzler, Ruggero Lazzaroni, Apeksha Shetty, and Jana
  Lasser.
\newblock Best practices for source-based research on misinformation and news
  trustworthiness using newsguard.
\newblock {\em Journal of Quantitative Description: Digital Media}, 5, January
  2025.

\bibitem{mcinnes_umap_2020}
Leland McInnes, John Healy, and James Melville.
\newblock {UMAP}: {Uniform} {Manifold} {Approximation} and {Projection} for
  {Dimension} {Reduction}, September 2020.
\newblock arXiv:1802.03426 [stat].

\bibitem{melchior_systematic_2024}
Cristiane Melchior and Mírian Oliveira.
\newblock A systematic literature review of the motivations to share fake news
  on social media platforms and how to fight them.
\newblock {\em New Media \& Society}, 26(2):1127--1150, February 2024.

\bibitem{mohr2023inference}
Sebastian~Bernd Mohr.
\newblock Inference of modular structures in dynamical systems and the
  application to telegram data.
\newblock Master's thesis, Georg-August-Universität Göttingen, 2023.

\bibitem{monti_fake_2019}
Federico Monti, Fabrizio Frasca, Davide Eynard, Damon Mannion, and Michael~M.
  Bronstein.
\newblock Fake {News} {Detection} on {Social} {Media} using {Geometric} {Deep}
  {Learning}, February 2019.
\newblock arXiv:1902.06673 [cs].

\bibitem{muennighoff_mteb_2023}
Niklas Muennighoff, Nouamane Tazi, Loïc Magne, and Nils Reimers.
\newblock {MTEB}: {Massive} {Text} {Embedding} {Benchmark}, March 2023.
\newblock arXiv:2210.07316 [cs].

\bibitem{langdetect}
Shuyo Nakatani.
\newblock Language detection library for java, 2010.

\bibitem{newman_reuters_2024}
Nic Newman, Richard Fletcher, Craig~T. Robertson, Amy Ross~Arguedas, and
  Rasmus~Kleis Nielsen.
\newblock Reuters {Institute} digital news report 2024.
\newblock Technical report, Reuters Institute for the Study of Journalism,
  2024.

\bibitem{nguyen_deep_2019}
Thanh~Thi Nguyen, Quoc Viet~Hung Nguyen, Dung~Tien Nguyen, Duc~Thanh Nguyen,
  Thien Huynh-The, Saeid Nahavandi, Thanh~Tam Nguyen, Quoc-Viet Pham, and
  Cuong~M. Nguyen.
\newblock Deep {Learning} for {Deepfakes} {Creation} and {Detection}: {A}
  {Survey}.
\newblock {\em arXiv preprint arXiv:1909.11573}, 2019.
\newblock Publisher: arXiv Version Number: 5.

\bibitem{Notarmuzi2022}
Daniele Notarmuzi, Claudio Castellano, Alessandro Flammini, Dario Mazzilli, and
  Filippo Radicchi.
\newblock Universality, criticality and complexity of information propagation
  in social media.
\newblock {\em Nature Communications}, 13(1), March 2022.

\bibitem{oxford-dictionary_word_2016}
Oxford-Dictionary.
\newblock Word of the year, 2016.

\bibitem{paszke2019pytorch}
Adam Paszke, Sam Gross, Francisco Massa, Adam Lerer, James Bradbury, Gregory
  Chanan, Trevor Killeen, Zeming Lin, Natalia Gimelshein, Luca Antiga, et~al.
\newblock Pytorch: An imperative style, high-performance deep learning library.
\newblock {\em Advances in neural information processing systems}, 32, 2019.

\bibitem{phan_fake_2023}
Huyen~Trang Phan, Ngoc~Thanh Nguyen, and Dosam Hwang.
\newblock Fake news detection: {A} survey of graph neural network methods.
\newblock {\em Applied Soft Computing}, 139:110235, May 2023.

\bibitem{pyg-team_torch_geometricnnconvgcnconv_2025}
PyG-Team.
\newblock torch\_geometric.nn.conv.{GCNConv} — pytorch\_geometric
  documentation, 2025.

\bibitem{Rieskamp2024}
Jonas Rieskamp, Milad Mirbabaie, Marie Langer, and Alexander Kocur.
\newblock From virality to veracity: Examining false information on telegram
  vs. twitter.
\newblock In {\em Proceedings of the 57th Hawaii International Conference on
  System Sciences}, HICSS. Hawaii International Conference on System Sciences,
  2024.

\bibitem{Saracco_2015}
Fabio Saracco, Riccardo Di~Clemente, Andrea Gabrielli, and Tiziano Squartini.
\newblock Randomizing bipartite networks: the case of the world trade web.
\newblock {\em Scientific Reports}, 5(1):10595, 2015.

\bibitem{Saracco_2017}
Fabio Saracco, Mika~J. Straka, Riccardo Di~Clemente, Andrea Gabrielli, Guido
  Caldarelli, and Tiziano Squartini.
\newblock Inferring monopartite projections of bipartite networks: an
  entropy-based approach.
\newblock {\em New Journal of Physics}, 19(5):053022, 2017.

\bibitem{BICM_Github}
Mika~J. Straka.
\newblock Bipartite configuration model for python.
\newblock \url{https://github.com/tsakim/bicm}, 2017.

\bibitem{vallarano_fast_2021}
Nicolò Vallarano, Matteo Bruno, Emiliano Marchese, Giuseppe Trapani, Fabio
  Saracco, Giulio Cimini, Mario Zanon, and Tiziano Squartini.
\newblock Fast and scalable likelihood maximization for {Exponential} {Random}
  {Graph} {Models} with local constraints.
\newblock {\em Scientific Reports}, 11(1):15227, July 2021.

\bibitem{varela_30_2023}
Layla Varela.
\newblock 30 {Interesting} {Telegram} {Statistics} you need to check (2025) -
  {Skillademia}, February 2023.
\newblock Section: Statistics.

\bibitem{velickovic_graph_2018}
Petar Veličković, Guillem Cucurull, Arantxa Casanova, Adriana Romero, Pietro
  Liò, and Yoshua Bengio.
\newblock Graph {Attention} {Networks}, February 2018.
\newblock arXiv:1710.10903.

\bibitem{wang_viral_2020}
Xiaohui Wang and Yunya Song.
\newblock Viral misinformation and echo chambers: the diffusion of rumors about
  genetically modified organisms on social media.
\newblock {\em Internet Research}, 30(5):1547--1564, June 2020.

\bibitem{Wehrli2025}
Silvan Wehrli, Anna-Maria Hartner, T~Sonia Boender, Bert Arnrich, and
  Christopher Irrgang.
\newblock Information pathways and voids in critical german online communities
  during the covid-19 vaccination discourse: Cross-platform and mixed methods
  analysis.
\newblock {\em Journal of Medical Internet Research}, 27:e76309–e76309,
  October 2025.

\bibitem{wijermars_is_2022}
Mariëlle Wijermars, , and Tetyana Lokot.
\newblock Is {Telegram} a “harbinger of freedom”? {The} performance,
  practices, and perception of platforms as political actors in authoritarian
  states.
\newblock {\em Post-Soviet Affairs}, 38(1-2):125--145, March 2022.
\newblock Publisher: Routledge \_eprint:
  https://doi.org/10.1080/1060586X.2022.2030645.

\bibitem{zhang_fakedetector_2020}
Jiawei Zhang, Bowen Dong, and Philip~S. Yu.
\newblock {FakeDetector}: {Effective} {Fake} {News} {Detection} with {Deep}
  {Diffusive} {Neural} {Network}.
\newblock In {\em 2020 {IEEE} 36th {International} {Conference} on {Data}
  {Engineering} ({ICDE})}, pages 1826--1829, April 2020.
\newblock ISSN: 2375-026X.

\bibitem{furht_social_2010}
Mingxin Zhang.
\newblock Social {Network} {Analysis}: {History}, {Concepts}, and {Research}.
\newblock In Borko Furht, editor, {\em Handbook of {Social} {Network}
  {Technologies} and {Applications}}, pages 3--21. Springer US, New York, NY,
  2010.

\bibitem{zhou2019network}
Xinyi Zhou and Reza Zafarani.
\newblock Network-based fake news detection: A pattern-driven approach.
\newblock {\em ACM SIGKDD explorations newsletter}, 21(2):48--60, 2019.

\end{thebibliography}
\end{document}